%% file: main.tex
\pdfoutput=1
\documentclass[letterpaper,twocolumn,10pt]{article}
\usepackage[T2A,T1]{fontenc}
\usepackage{usenix}
\usepackage{cite}
\usepackage{amsmath,amssymb,amsfonts}
\usepackage{algorithmic}
\usepackage{graphicx}
\usepackage{textcomp}
\usepackage{url}

\usepackage{hyperref}
\usepackage{cleveref}
\usepackage{booktabs}
\usepackage{xspace}
\usepackage{caption}
\usepackage{enumitem}
\usepackage{listings}
\usepackage{multirow}
\usepackage{fancyhdr}

\hypersetup{
    pdftitle={Trojan Source: Invisible Vulnerabilities},
    pdfauthor={Boucher and Anderson}
}

\fancypagestyle{firststyle}
{
   \fancyhf{}
   \chead{\raisebox{0.5in}[0in][0in]{\textit{--------------------------------------- In 32nd USENIX Security Symposium ---------------------------------------}}}
}

\begin{document}

\newcommand{\cyrillicx}{{\fontencoding{T2A}\selectfont х\xspace}}
\newcommand{\cyrillicH}{{\fontencoding{T2A}\selectfont Н\xspace}}
\lstset{basicstyle=\small\ttfamily}

\title{Trojan Source:\\ Invisible Vulnerabilities}

\author{
{\rm Nicholas Boucher}\\
\small{University of Cambridge}\\
\small{Computer Science \& Technology}\\
\small{\rm{}nicholas.boucher@cl.cam.ac.uk}
\and
{\rm Ross Anderson}\\
\small{University of Cambridge}\\
\small{and University of Edinburgh}\\
\small{\rm{}ross.anderson@cl.cam.ac.uk}
}

\maketitle

\thispagestyle{firststyle}

\input{sections/abstract}

\input{sections/introduction}
\input{sections/background}
\input{sections/methodology}
\input{sections/exploitation}
\input{sections/related}
\input{sections/evaluation}
\input{sections/discussion}
\input{sections/conclusion}

\section*{Acknowledgment}

We would like to thank GitHub for assisting with scanning the open source ecosystem, Pietro Albini and Mara Bos of the Rust project for scanning crates.io, CERT/CC for assistance with coordinated disclosure, GitHub user @security-curious for suggesting Bidi control characters in identifiers and RegExs, and Abdun Nihaal for suggesting invisible characters in multiline comments.

\bibliographystyle{IEEEtran}
{\small
\bibliography{sources}}

\clearpage
\onecolumn
\appendix
\input{sections/appendix}

\end{document}

%% file: sections/abstract.tex
\begin{abstract}
We present a new type of attack in which source code is maliciously encoded so that it appears different to a compiler and to the human eye. This attack exploits subtleties in text-encoding standards such as Unicode to produce source code whose tokens are logically encoded in a different order from the one in which they are displayed, leading to vulnerabilities that cannot be perceived directly by human code reviewers. `Trojan Source' attacks, as we call them, pose an immediate threat both to first-party software and of supply-chain compromise across the industry. We present working examples of Trojan Source attacks in C, C++, C\#, JavaScript, Java, Rust, Go, Python, SQL, Bash, Assembly, and Solidity. We propose definitive compiler-level defenses, and describe other mitigating controls that can be deployed in editors, repositories, and build pipelines while compilers are upgraded to block this attack. We document an industry-wide coordinated disclosure for these vulnerabilities; as they affect most compilers, editors, and repositories, the exercise teaches how different firms, open-source communities, and other stakeholders respond to vulnerability disclosure.
\end{abstract}

%% file: sections/introduction.tex
\section{Introduction}

What if it were possible to trick compilers into emitting binaries that did not match the logic visible in source code? We demonstrate that this is not only possible for a broad class of modern compilers, but easily exploitable.

We show that subtleties of modern expressive text encodings, such as Unicode, can be used to craft source code that appears visually different to developers and to compilers. The difference can be exploited to invisibly alter the logic in an application and introduce targeted vulnerabilities.

The belief that trustworthy compilers emit binaries correctly implementing the algorithms defined in source code is a foundational assumption of software. It is well-known that malicious compilers can produce binaries containing vulnerabilities~\cite{DBLP:journals/cacm/Thompson84}; as a result, there has been significant effort devoted to verifying compilers and mitigating their exploitable side-effects. However, to our knowledge, producing vulnerable binaries via unmodified compilers by manipulating the encoding of otherwise non-malicious source code has not so far been explored. 

Consider a supply-chain attacker who seeks to inject vulnerabilities into software upstream of the ultimate targets, as happened in the recent Solar Winds incident~\cite{9382367}. Two methods an adversary may use to accomplish such a goal are suborning an insider to commit vulnerable code into software systems, and contributing subtle vulnerabilities into open-source projects. In order to prevent or mitigate such attacks, it is essential for developers to perform at least one code or security review of every submitted contribution. However, this critical control may be bypassed if the vulnerabilities do not appear in the source code displayed to the reviewer, but are hidden in the encoding layer underneath. Such an attack is quite feasible, as we will now demonstrate.

\input{tables/bidi-characters}

In this paper, we make the following contributions:
\begin{itemize}[itemsep=0em,topsep=1em]
    \item We define a novel class of vulnerabilities, which we call Trojan Source attacks, and which use maliciously encoded but semantically permissible source code modifications to introduce invisible software vulnerabilities.
    \item We provide working examples of Trojan Source vulnerabilities in C, C++, C\#, JavaScript, Java, Rust, Go, Python, SQL, Bash, Assembly, and Solidity.
    \item We describe effective defenses that must be employed by compilers, as well as other defenses that can be used in editors, repositories, and build pipelines, and discuss the limitations of these defenses.
    \item We document the coordinated disclosure process we used to disclose this vulnerability across the industry, and what it teaches about the response to disclosure.
    \item We raise a new question about what it means for a compiler to be trustworthy.
\end{itemize}

%% file: tables/bidi-characters.tex
\begin{table*}[t]
\centering
\caption{Unicode directionality formatting characters relevant to reordering attacks.\\See Bidi specification for complete list~\cite{unicode_bidi_2020}.}
\label{tab:bidi-overrides}
\resizebox{\textwidth}{!}{%
\begin{tabular}{llll}
\hline
Abbreviation & Code Point & Name & Description \\ \hline
LRE & \cellcolor[HTML]{FFFFFF}\texttt{U+202A} & Left-to-Right Embedding & Try treating following text as left-to-right. \\
RLE & \texttt{U+202B} & Right-to-Left Embedding & Try treating following text as right-to-left. \\
LRO & \texttt{U+202D} & Left-to-Right Override & Force treating following text as left-to-right. \\
RLO & \texttt{U+202E} & Right-to-Left Override & Force treating following text as right-to-left. \\
LRI & \texttt{U+2066} & Left-to-Right Isolate & Force treating following text as left-to-right without affecting adjacent text. \\
RLI & \texttt{U+2067} & Right-to-Left Isolate & Force treating following text as right-to-left without affecting adjacent text. \\
FSI & \texttt{U+2068} & First Strong Isolate & Force treating following text in direction indicated by the next character. \\
PDF & \texttt{U+202C} & Pop Directional Formatting & Terminate nearest LRE, RLE, LRO, or RLO. \\
PDI & \texttt{U+2069} & Pop Directional Isolate & Terminate nearest LRI or RLI. \\ \hline
\end{tabular}%
}
\end{table*}

%% file: sections/background.tex
\section{Background}

\subsection{Compiler Security}

Compilers translate high-level programming languages into lower-level representations such as architecture-specific machine instructions or portable bytecode. They seek to implement the formal specifications of their input languages, deviations from which are considered to be bugs.

Since the 1960s~\cite{painter_correctness_1967}, researchers have investigated formal methods to mathematically prove that a compiler's output correctly implements the source code supplied to it~\cite{dave2003compiler,patterson2019next}. Many of the discrepancies between source code logic and compiler output logic stem from compiler optimizations, about which it can be difficult to reason~\cite{7163211}. These optimizations may also cause side-effects that have security consequences~\cite{8406587}.

\subsection{Text Encodings}

Digital text is stored as an encoded sequence of numerical values, or code points, that correspond with characters according to the relevant specification. While single-script specifications such as ASCII were historically prevalent, modern text encodings have standardized\footnote{According to scans by \href{https://w3techs.com/technologies/details/en-utf8}{w3techs.com/technologies/details/en-utf8}, 97\% of the most accessed 10 million websites in 2021 use UTF-8 Unicode encodings.} around Unicode~\cite{unicode_2020}.

At the time of writing, Unicode defines 143,859 characters across 154 different scripts in addition to various non-script character sets (such as emojis) plus a plethora of control characters. While its specification provides a mapping from numerical code points to characters, the binary representation of those code points is determined by which of various encodings is used, with one of the most common being UTF-8.

Text rendering is performed by interpreting encoded bytes as numerical code points according to the chosen encoding, then looking up the characters in the relevant specification, then resolving all control characters, and finally displaying the glyphs provided for each character in the chosen font.

\subsection{Supply-Chain Attacks}

Supply-chain attacks are those in which an adversary tries to introduce targeted vulnerabilities into deployed applications, operating systems, and software components~\cite{5718996}. Once published, such vulnerabilities are likely to persist within the affected ecosystem even if patches are later released~\cite{7163055}. Following a number of attacks that compromised multiple firms and government departments, supply-chain attacks have gained urgent attention from the US White House~\cite{jospeh_biden_jr_executive_2021}.

Adversaries may introduce vulnerabilities in supply-chain attacks through modifying source code, compromising build systems, or attacking the distribution of published software~\cite{5428501,1203227}. Distribution attacks are mitigated by software producers signing binaries, so attacks on the earlier stages of the pipeline are particularly attractive. Attacks on upstream software such as widely-utilized packages can affect multiple dependent products, potentially compromising whole ecosystems. As supply-chain threats involve multiple organizations, modeling and mitigating them requires consideration of technical, economic, and social factors~\cite{7180277}.

Open-source software provides a significant vector through which supply-chain attacks can be launched~\cite{10.1007/978-3-030-52683-2_2}, and is ranked as one of OWASP's Top 10 web application security risks~\cite{owasp_a92017}.

%% file: sections/methodology.tex
\section{Attack Methodology}

\subsection{Reordering}

Internationalized text encodings require support for both left-to-right languages such as English and Russian, and right-to-left languages such as Hebrew and Arabic. When mixing scripts with different display orders, there must be a deterministic way to resolve conflicting directionality. For Unicode, this is implemented in the Bidirectional, or Bidi, Algorithm~\cite{unicode_bidi_2020}.

In some scenarios, the default ordering set by the Bidi Algorithm may not be sufficient; for these cases, Bidi control characters are provided. Bidi control characters are invisible characters that enable switching the display ordering of groups of characters.

\Cref{tab:bidi-overrides} provides a list of Bidi control characters relevant to this attack. Of note are \texttt{LRI} and \texttt{RLI}, which format subsequent text as left-to-right and right-to-left respectively, and are both closed by \texttt{PDI}.

Bidi control characters enable even single-script characters to be displayed in an order different from their logical encoding. This fact has previously been exploited to disguise the file extensions of malware disseminated by email~\cite{brian_krebs_right--left_2011} and to craft adversarial examples for NLP machine-learning pipelines~\cite{boucher2021bad}.

As an example, consider the following Unicode character sequence:
$$\texttt{{\color{blue}RLI} a b c {\color{blue}PDI}}$$
which will be displayed as:
$$\texttt{c b a}$$

All Unicode Bidi control characters are restricted to affecting a single paragraph, as a newline character will explicitly close any unbalanced control characters, namely those that lack a corresponding closing character.

\input{figures/py-2}

\subsection{Isolate Shuffling}

In the Bidi specification, isolates are groups of characters that are treated as a single entity; that is, the entire isolate will be moved as a single block when the display order is overridden.

Isolates can be nested. For example, consider the Unicode character sequence:
$$\texttt{{\color{blue}RLI} {\color{red}LRI} a b c {\color{red}PDI} {\color{PineGreen}LRI} d e f {\color{PineGreen}PDI} {\color{blue}PDI}}$$
which will be displayed as:
$$\texttt{d e f a b c}$$

Embedding multiple layers of \texttt{LRI} and \texttt{RLI} within each other enables the near-arbitrary reordering of strings. This gives an adversary fine-grained control, so they can manipulate the display order of text into an anagram of its logically-encoded order.

\subsection{Compiler Manipulation}

Like most non-text rendering systems, compilers and interpreters do not typically process formatting control characters, including Bidi control characters, prior to parsing source code. This can be used to engineer a targeted gap between the visually-rendered source code as seen by a human eye, and the raw bytes of the encoded source code as evaluated by a compiler.

We can exploit this gap to create adversarially-encoded text that is understood differently by human reviewers and by compilers.

\subsection{Syntax Adherence}

Most well-designed programming languages will not allow arbitrary control characters in source code, as they will be viewed as tokens meant to affect the logic. Thus, randomly placing Bidi control characters in source code will typically result in a compiler or interpreter syntax error. To avoid such errors, we can exploit two general principles of programming languages:

\begin{itemize}
    \item \textbf{Comments} -- Most programming languages allow comments within which all text (including control characters) is ignored by compilers and interpreters.
    \item \textbf{Strings} -- Most programming languages allow string literals that may contain arbitrary characters, including control characters.
\end{itemize}

While both comments and strings will have syntax-specific semantics indicating their start and end, these bounds are not respected by Bidi control characters. Therefore, by placing Bidi control characters exclusively within comments and strings, we can smuggle them into source code in a manner that most compilers will accept.

Making a random modification to the display order of characters on a line of valid source code is not particularly interesting, as it is very likely to be noticed by a human reviewer. Our key insight is that we can reorder source code characters in such a way that the resulting display order also represents syntactically valid source code.

\subsection{Novel Supply-Chain Attack}

Bringing all this together, we arrive at a novel supply-chain attack on source code. By injecting Unicode Bidi control characters into comments and strings, an adversary can produce syntactically-valid source code in most modern languages for which the display order of characters presents logic that diverges from the real logic. In effect, we anagram program A into program B.

Such an attack could be challenging for a human code reviewer to detect, as the rendered source code looks perfectly acceptable. If the change in logic is subtle enough to go undetected in subsequent testing, an adversary could introduce targeted vulnerabilities without being detected. We provide working examples of this attack in the following section.

Yet more concerning is the fact that Bidi control characters persist through the copy-and-paste functions on most modern browsers, editors, and operating systems. Any developer who copies code from an untrusted source into a protected code base may inadvertently introduce an invisible vulnerability. Code copying is already a significant source of real-world security exploits~\cite{7546508}.

\subsection{Threat Model}

More formally, we define the threat model for Trojan Source attacks as an active adversary who seeks to inject adversarial logic into targeted software. If such software has an upstream dependency on further software, the adversary may target that instead in a supply chain attack. We define the adversary as having the following access:
\begin{itemize}[itemsep=0em,topsep=.5em]
    \item write access to that target software's source code, such as via a direct pull request, \textit{or}
    \item write access to an upstream dependency of the target software, such as via a pull request against an open source project, \textit{or}
    \item the ability to post code samples that will be copied and pasted into the target software's source code, such as via question answering websites~\cite{stackoverflow_android,stackoverflow_voting}.
\end{itemize}

\subsection{Generality}

We have implemented the above attack methodology and the examples in the following section with Unicode. Many modern compilers accept Unicode source code, as will be noted in our experimental evaluation. However, this attack paradigm should work with any text specification that enables the manipulation of display order, which is necessary to support internationalized text.

Should the Unicode specification be supplanted by another standard, then in the absence of specific defenses, we believe that it is very likely to provide the same bidirectional functionality used to perform this attack. To substantiate this conjecture, we repeated the experiments presented throughout this paper using Chinese standard GB18030 and Israeli standard SI1311:2002 in addition to UTF-8 achieving the same results across all three specifications.

%% file: figures/py-2.tex
\begin{figure*}[t]
  \centering
  \begin{minipage}[b]{0.49\textwidth}
    \centering
    \frame{\includegraphics[width=\linewidth]{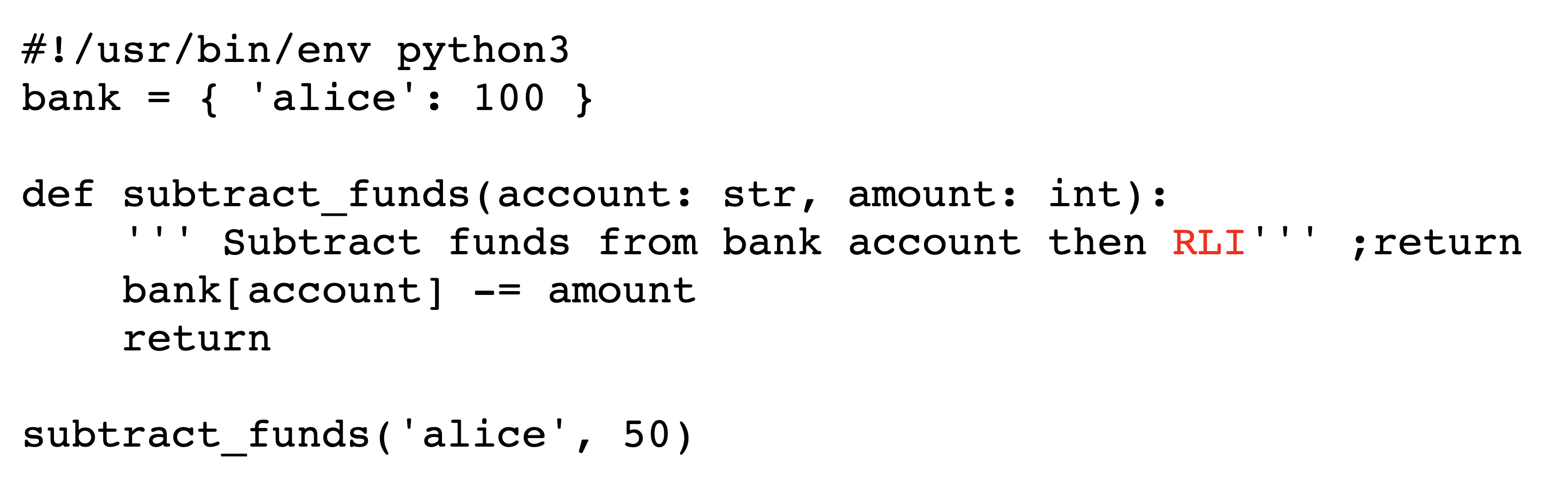}}
    \caption{Encoded bytes of a Trojan Source early-return attack in Python.}
    \label{fig:py-2-encoded}
  \end{minipage}
  \hfill
  \begin{minipage}[b]{0.49\textwidth}
    \centering
    \frame{\includegraphics[width=\linewidth]{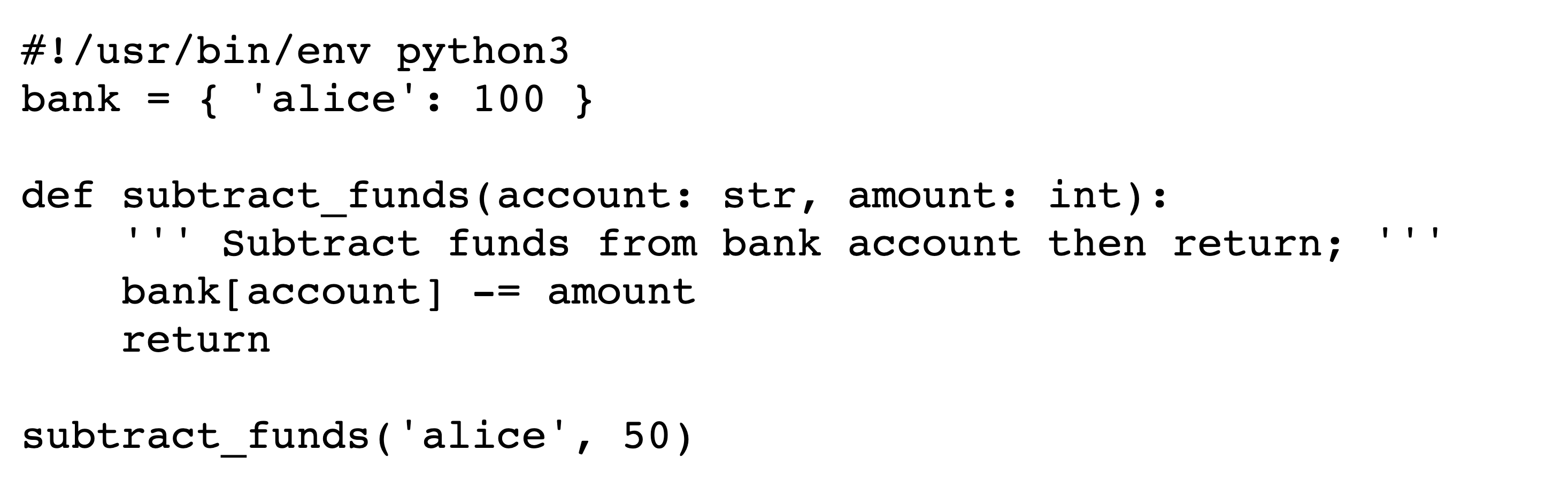}}
    \caption{Rendered text of a Trojan Source early-return attack in Python.}
    \label{fig:py-2-rendered}
  \end{minipage}
\end{figure*}

%% file: sections/exploitation.tex
\section{Exploit Techniques}

\input{figures/c-2}

There are a variety of ways to exploit the adversarial encoding of source code. The underlying principle is the same in each: use Bidi control characters to create a syntactically valid reordering of source code characters in the target language.

In the following section, we propose three general types of exploits that work across multiple languages. We do not claim that this list is exhaustive.

\subsection{Early Returns}

In the early-return exploit technique, adversaries disguise a genuine return statement as a comment or string literal, so they can cause a function to return earlier than it appears to.

Consider, for example, the case of docstrings -- formal comments that purport to document the purpose of a function -- which are considered good practice in software development. In languages where docstrings can be located within a function definition, an adversary need only find a plausible location to write the word \texttt{return} (or its language-specific equivalent) in a docstring comment, and then reorder the comment such that the \texttt{return} statement is executed immediately following the comment.

\Cref{fig:py-2-encoded,fig:py-2-rendered} depict the encoded bytes and rendered text, respectively, of an early-return attack in Python3. Viewing the rendered text of the source code in~\Cref{fig:py-2-rendered}, one would expect the value of \texttt{bank[\textquotesingle alice\textquotesingle]} to be 
\texttt{50} after program execution. However, the value of \texttt{bank[\textquotesingle alice\textquotesingle]} remains \texttt{100} after the program executes. This is because the word \texttt{return} in the docstring is actually executed due to a Bidi control character, causing the function to return prematurely and the code which subtracts value from a user's bank account to never run.

This technique is not specific to docstrings; any comment or string literal that can be manipulated by an adversary could hide an early-return statement.

\subsection{Commenting-Out}

In this exploit technique, text that appears to be legitimate code actually exists within a comment and is thus never executed. This allows an adversary to show a reviewer some code that appears to be executed but is not present from the perspective of the compiler or interpreter. For example, an adversary can comment out an important conditional, and then use Bidi control characters to make it appear to be still present.

This method is easiest to implement in languages that support mutliline comments. An adversary begins a line of code with a multiline comment that includes the code to be commented out and closes the comment on the same line. They then need only insert Bidi control characters to make it appear as if the comment is closed before the code via isolate shuffling.

\Cref{fig:c-2-encoded,fig:c-2-rendered} depict the encoded bytes and rendered text, respectively, of a commenting-out attack in C. Viewing the rendered text makes it appear that, since the user is not an admin, no text should be printed. However, upon execution the program prints \textit{You are an admin.} The conditional does not actually exist; in the logical encoding, its text is wholly within the comment.

The previous example is aided by the Unicode feature that directionality-aware punctuation characters are displayed in reverse within right-to-left settings, e.g. \texttt{\{} becomes \texttt{\}}. This can be particularly insidious for the following symbols typically used for inequality tests and bit shifts: \texttt{<{}<}, \texttt{>{}>}, \texttt{<}, and \texttt{>}.

\subsection{Stretched Strings}

In this exploit technique, text that appears to be outside a string literal is actually located within it. This allows an adversary to manipulate string comparisons, for example causing strings which appear identical to give rise to a non-equal comparison.

\Cref{fig:js-1-encoded,fig:js-1-rendered} depict the encoded bytes and rendered text, respectively, of a stretched-string attack in JavaScript. While it appears that the user's access level is \textquotedbl user\textquotedbl\ and therefore nothing should be written to the console, the code in fact outputs \textit{You are an admin.} This is because the apparent comment following the comparison isn't actually a comment, but included in the comparison's string literal.

In general, the stretched-strings technique will allow an adversary to cause string comparisons to fail. In languages that support a limited set of alternate literals, such as regular expression literals in JavaScript, the stretched string technique can be generalized to apply. A small set of languages, such as Ruby, support Bidi control characters in identifiers such as variable names, and in these languages this technique also generalizes.

However, there are other, perhaps simpler, ways that an adversary can cause a string comparison to fail without visual effect. For example, the adversary can place invisible characters -- that is, characters in Unicode that render to the absence of a glyph -- such as the Zero Width Space\footnote{Unicode character \texttt{U+200B}} (ZWSP) into string literals used in comparisons. Although these invisible characters do not change the way a string literal renders, they will cause string comparisons to fail. Another option is to use characters that look the same, known as homoglyphs, such as the Cyrillic letter `{\fontfamily{Libertine}\selectfont\cyrillicx}' which typically renders identical to the Latin letter `{\fontfamily{Libertine}\selectfont x}' used in English but occupies a different code point. Depending on the context, the use of other character-encoding tricks may be more desirable than a stretched-string attack using Bidi control characters.

\input{figures/js-1}

%% file: figures/c-2.tex
\begin{figure*}[t]
  \centering
  \begin{minipage}[b]{0.49\textwidth}
    \centering
    \frame{\includegraphics[width=\linewidth]{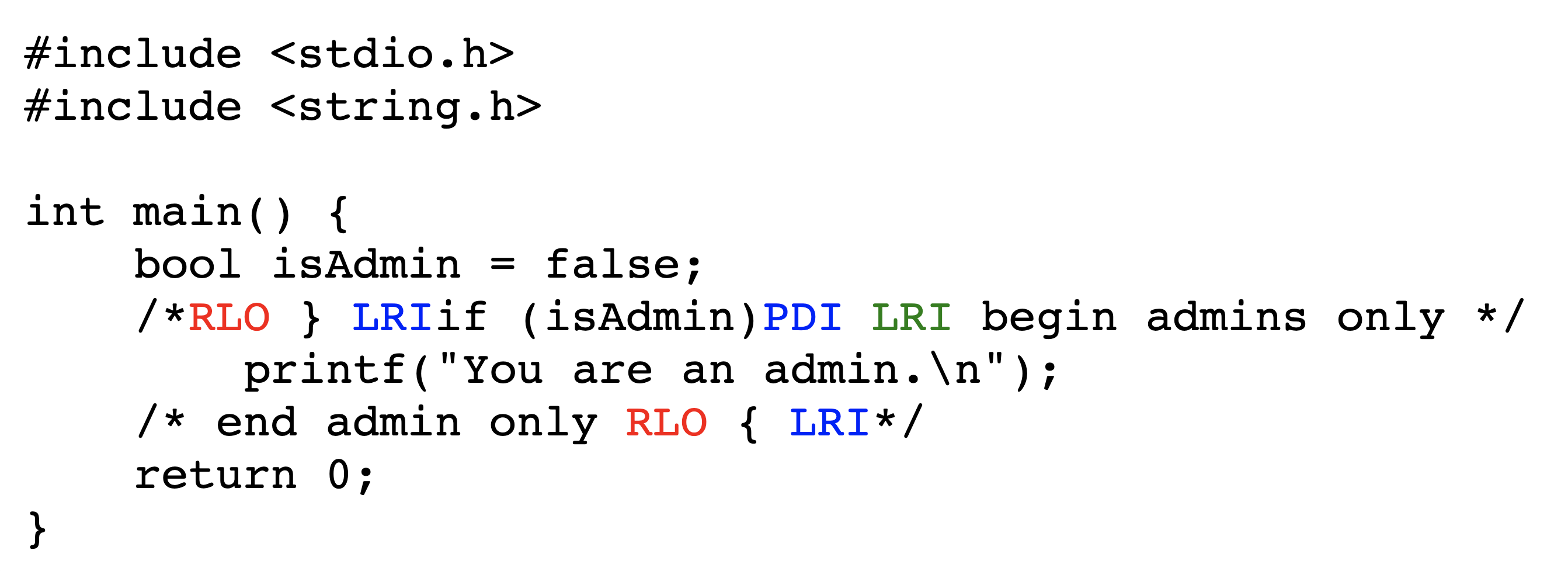}}
    \caption{Encoded bytes of a Trojan Source commenting-out attack in C.}
    \label{fig:c-2-encoded}
  \end{minipage}
  \hfill
  \begin{minipage}[b]{0.49\textwidth}
    \centering
    \frame{\includegraphics[width=\linewidth]{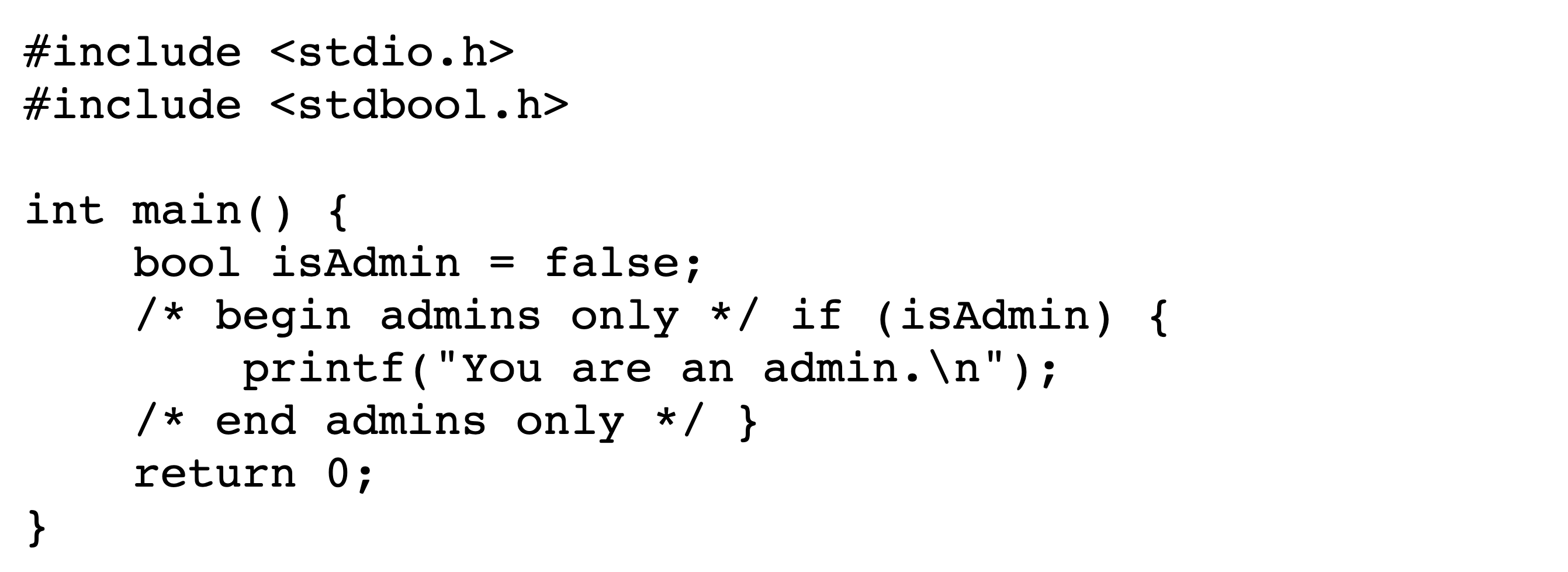}}
    \caption{Rendered text of a Trojan Source commenting-out attack in C.}
    \label{fig:c-2-rendered}
  \end{minipage}
\end{figure*}

%% file: figures/js-1.tex
\begin{figure*}[t]
  \centering
  \begin{minipage}[b]{0.49\textwidth}
    \centering
    \frame{\includegraphics[width=\linewidth]{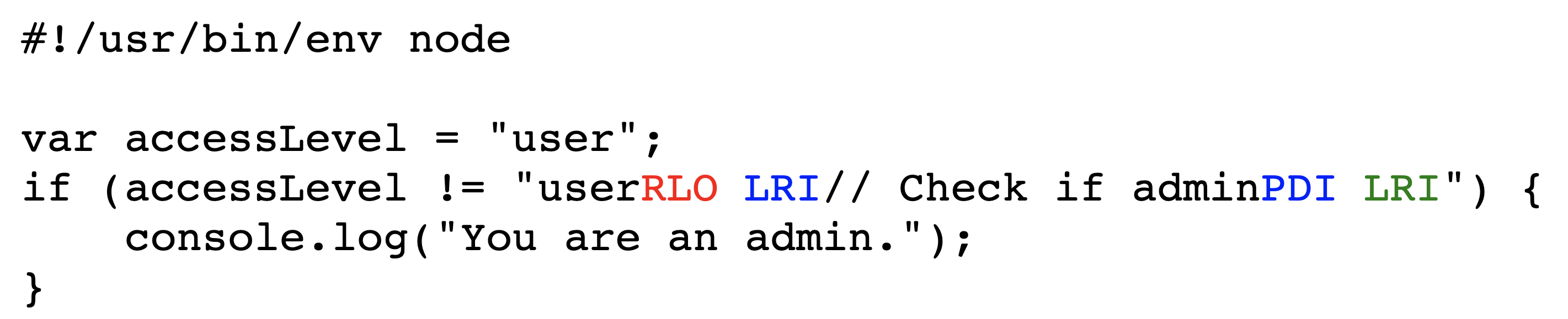}}
    \caption{Encoded bytes of a Trojan Source stretched-string attack in JavaScript.}
    \label{fig:js-1-encoded}
  \end{minipage}
  \hfill
  \begin{minipage}[b]{0.49\textwidth}
    \centering
    \frame{\includegraphics[width=\linewidth]{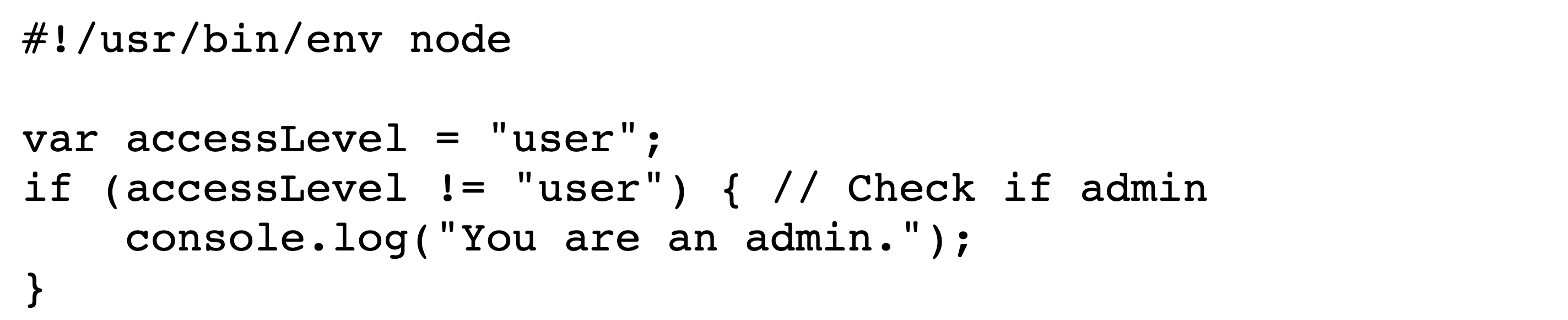}}
    \caption{Rendered text of a Trojan Source stretched-string attack in JavaScript.}
    \label{fig:js-1-rendered}
  \end{minipage}
\end{figure*}

%% file: sections/related.tex
\section{Related Work}

\subsection{URL Security}

Deceptively encoded URLs have long been a tool of choice for spammers~\cite{SimpsonMC20apwg}, with one of the earliest documented examples being the case of \textit{paypaI.com}. This July 2000 campaign sought to trick users into disclosing passwords for \textit{paypal.com} by registering a domain with the lowercase l replaced  with the visually similar uppercase I~\cite{sullivan_paypal_2000}.

These domain attacks become even more severe with the introduction of Unicode, which has a much larger set of visually similar characters, or homoglyphs, than ASCII. In fact, Unicode produces a security report which spends considerable length discussing domain-related concerns~\cite{unicode_security_2014}, and the topic of homoglyphs in URLs has been thoroughly examined in the literature~\cite{10.1145/503124.503156,10.5555/1267359.1267383,noauthor_capec-632_2015,10.1145/3355369.3355587}.

Punycode~\cite{RFC3492}, a standard for converting Unicode URLs to ASCII, bootstraps DNS support for internationalized domains but does not mitigate homoglyph attacks. It is used in conjunction with IDNA~\cite{RFC5891}, which sets rules for handling Bidi and invisible characters to prevent some look-alike domains.

\subsection{Adversarial NLP}

Bidi control characters and homoglyphs have both been used to create adversarial examples in the machine learning NLP setting~\cite{boucher2021bad}. These characters, together with invisible characters such as zero-width spaces and deletions control characters, are used to generate strings that look visually identical to some target string but are represented by different Unicode encodings. Optimal encodings are discovered using a gradient-free optimization method that can be used to manipulate the performance of models in both a targeted and untargeted fashion.

\subsection{Visually Deceptive Malware}

Bidi overrides have historically been used in the wild to change the appearance of file extensions~\cite{brian_krebs_right--left_2011}. This technique aids email-based distribution of malware, as it can deceive a user into running an executable file when they believe they are opening something more benign. Similarly, directionality control characters have been used in at least one family of malware to disguise the names of malicious system services~\cite{microsoft_win32sirefef_2017}.

Attacks have also been proposed in which an adversary uses homoglyphs to create filenames that look visually similar to key system files, and then replaces references to those files with the adversarial homoglyph version~\cite{jakob_lell_hacking-contest_2014}.

In general, purposefully confusing code written to obscure vulnerabilities is known as \textit{underhanded source code}, and a series of competitions have been historically held to evaluate proposed underhanded methods~\cite{wheeler_2020_underhanded}. Many vulnerability patterns fall under this heading. Common techniques used in underhanded source code competitions include replacing numbers with letters, leveraging out-of-bounds reads and writes, swapping equality and assignment operators, and misusing macros. Trojan Source attacks could be considered to belong to this class of vulnerability patterns.

\subsection{Software Vulnerabilities}

In addition to purposefully crafted malware, attackers can exploit common vulnerabilities in otherwise benign software to introduce adversarial behavior~\cite{nistvulntaxonomy}. When discovered, vulnerabilities are tracked under common identifiers known as CVEs~\cite{mitre_cve_2021}. These vulnerabilities may be hard to detect when viewing source code, such as in the case of return oriented programming~\cite{solardesigner1997,10.1145/2133375.2133377} which abuses return statements to execute assembly instructions in an unexpected order. Trojan Source attacks could also be considered to belong to this class.

%% file: sections/evaluation.tex
\section{Evaluation}

\subsection{Experimental Setup}

To validate the feasibility of the attacks described in this paper, we implemented proof-of-concept attacks on simple programs in 12 different languages. Each proof of concept is a program with source code that, when rendered, displays logic indicating that the program should have no output; however, the compiled version of each program outputs the text `\textit{You are an admin.}' due to Trojan Source attacks using Bidi control character encodings.

For this attack paradigm to work, the compilers or interpreters used must accept some form of Unicode input, such as UTF-8. We find that this is true for the overwhelming majority of languages in modern use. It is also necessary for the language to syntactically support modern internationalized text in string literals or comments.

Thanks to our disclosure process, compilers and interpreters are starting to employ defenses that emit errors or warnings when this attack is detected, as are some editors, but we found no evidence of such behavior in any of the experiments we conducted before starting the process. At the time of writing, none of the language specifications have been changed to prevent Trojan Source attacks. We discuss the results of the disclosure process later.

All proofs of concept referenced in this paper have been made available online\footnote{\href{https://github.com/nickboucher/trojan-source}{github.com/nickboucher/trojan-source}}. We have also created a website to help disseminate knowledge of this vulnerability pattern to all developer communities\footnote{\href{https://trojansource.codes}{trojansource.codes}}.

\input{tables/languages}

\subsection{Languages}

The following sections describe and evaluate Trojan Source attack proofs-of-concept against specific programming languages. The results are presented in \Cref{tab:langs}.

\subsubsection{C}

As previously discussed, \Cref{fig:c-2-rendered,fig:c-2-encoded} depict a commenting-out attack in C. We also provide an example of a Stretched-String attack in C in \Cref{apx:c}.

In addition to supporting string literals, C supports both single-line and multi-line comments~\cite{ISO:2018:III}. Single-line comments begin with the sequence \texttt{//} and are terminated by a newline character. Multi-line comments begin with the sequence \texttt{$/^*$} and are terminated with the sequence \texttt{$^*/$}. Conveniently, multi-line comments can begin and end on a single line, despite their name. String literals are contained within double quotes, e.g. \textquotedbl~$\cdot$~\textquotedbl. Strings can be compared using the function \texttt{strcmp}.

C is well-suited for the commenting-out and stretched-string exploit techniques, but only partially suited for early returns. This is because when the multiline comment terminator, i.e. \texttt{*/}, is reordered using a right-to-left control character, it becomes \texttt{/*}. This provides a visual clue that something is not right. This can be overcome by writing reversible comment terminators as \texttt{/*/}, but this is less elegant and still leaves other visual clues such as the line-terminating semicolon. We provide an example of a functioning but less elegant early-return attack in C in \Cref{apx:c} 
which, although it looks like it prints `\textit{Hello World.}', in fact prints nothing.


\subsubsection{C++}

Since C++ is a linguistic derivative of C, it should be no surprise that the same attack paradigms work against the C++ specification~\cite{ISO:2020:IIIa}. Similar proofs-of-concept modified to adhere to C++ preferred syntax can be seen in \Cref{apx:c++}.


\subsubsection{C\#}

C\# is an object-oriented language created by Microsoft that typically runs atop .NET, a cross-platform managed runtime, and is used heavily in corporate settings~\cite{ISO:2018:CS}. C\# is vulnerable to the same attack paradigms as the preceding languages, and we present the same proof-of-concept attacks using C\# syntax in our online repository.

To our surprise, we found that C\# allows Bidi control characters in identifiers such as variable names. Even more surprisingly, these control characters can be placed arbitrarily within identifiers without affect. That is, the same variable can be referenced with or without a Bidi control character and it will resolve the same.


\footnotetext{All languages depicted are vulnerable; for specific attack techniques, \checkmark means the rendered code visually matches common style for that language, while \textasciitilde\ means visual renderings adhere to language syntax but deviate from common style (e.g. the multiline comment terminator \texttt{*/} is written as \texttt{/*/}). Code samples in the Appendix provide explicit examples.}

\input{tables/viewer-evaluation}

\subsubsection{JavaScript}

JavaScript, also known as ECMAScript, is an interpreted language that provides in-browser client-side scripting for web pages, and is increasingly also used for server-side web application and API implementations~\cite{Ecma:2021}. JavaScript is vulnerable to the same attack paradigms as the preceding languages, and we present the same proof-of-concept attacks using JavaScript syntax in \Cref{apx:js} 
as well as the previously discussed \Cref{fig:js-1-encoded,fig:js-1-rendered}.


\subsubsection{Java}

Java is a bytecode-compiled multipurpose language maintained by Oracle~\cite{gosling_java_2021}. It too is vulnerable to the same attack paradigms as the preceding languages, and we present the same proofs-of-concept using Java syntax in \Cref{apx:java}.


\subsubsection{Rust}

Rust is a high-performance language increasingly used in systems programming~\cite{rustref}. It too is vulnerable to the same attack paradigms as the preceding languages, and we present the same proof-of-concept attacks using Rust syntax in \Cref{apx:rust}. 
We note that the commenting-out attack throws an unused variable warning, but this is trivially avoidable.


\subsubsection{Go}

Go is a multipurpose open-source language produced by Google~\cite{goref}. Go is vulnerable to the same attack paradigms as as the preceding languages, and we present the same proof-of-concept attacks using Go syntax in \Cref{apx:go} and our online repository.


\subsubsection{Python}

Python is a general-purpose scripting language used heavily in data science and many other settings~\cite{pythonref}. Python supports multiline comments in the form of docstrings opened and closed with \texttt{\textquotesingle\textquotesingle\textquotesingle} or \texttt{\textquotedbl\textquotedbl\textquotedbl}. We have already exploited this fact in \Cref{fig:py-2-encoded,fig:py-2-rendered} to craft early-return attacks.

An additional commenting-out proof-of-concept attack against Python 3 can be found in encoded form in \Cref{apx:py}.


\subsubsection{SQL}

SQL is a common query language supporting optionally terminated C-style multiline comments. SQL is vulnerable to each attack technique as we demonstrate in our online proofs-of-concept repository.

\subsubsection{Bash}

Bash is a common shell script also vulnerable to each attack technique as demonstrated in our online proofs-of-concept repository.

\subsubsection{Assembly}

Assembly is a human-readable representation of machine instructions. Despite being a low-level language, it permits comments and string literals making it vulnerable to each attack technique as demonstrated in our online proofs-of-concept repository.

\subsubsection{Solidity}

Solidity is a language used to author smart contracts for the Etherium blockchain. Of all languages considered, Solidity is the only one that had partial compiler defenses in place against Bidi control characters prior to coordinated disclosure. The solidity compiler throws an error when Bidi override and embedding control characters are detected in source code; however, no errors are thrown for Bidi isolate control characters, rendering the defenses ineffective. We demonstrate this in our online proofs-of-concept repository.

\subsection{Code Viewers}

We were curious to see how these attacks were visualized by the editors and code repository front-ends used in modern development environments, as many tools have different Unicode implementations. We therefore tested the latest releases of the Visual Studio Code, Atom, Sublime Text, Notepad++, Eclipse, IntelliJ, vim, and emacs code editors as of October 2021. We also tested the GitHub, Bitbucket, and GitLab web-based code repository front-end interfaces as the same time. Each evaluation was repeated across three machines running Windows 10, MacOS Big Sur, and Ubuntu 20.04. The results can be found in \Cref{tab:viewers}, where \checkmark\ represents code that displayed the same as the example visualizations in this paper prior to coordinated disclosure. Applications that have since been patched are shaded. Any deviations are described.

%% file: tables/languages.tex
\begin{table*}[t]
\centering
\captionsetup{justification=centering}
\caption{Trojan Source attack language vulnerability.\\\checkmark represents fully vulnerable, and \textasciitilde\ represents vulnerable with less common style.\protect\footnotemark}
\label{tab:langs}
\begin{tabular}{lcccl}
\toprule
\textbf{Language} & \multicolumn{1}{l}{\textbf{Vulnerable}} & \multicolumn{1}{l}{\textbf{}} & \multicolumn{1}{l}{\textbf{}} & \textbf{Tool Evaluated} \\
\textbf{} & \multicolumn{1}{l}{Early Return} & \multicolumn{1}{l}{Commenting-Out} & \multicolumn{1}{l}{Stretched Strings} &  \\ \midrule
\rowcolor[HTML]{EFEFEF} 
\textbf{C} & \textasciitilde & \checkmark & \checkmark & \begin{tabular}[c]{@{}l@{}}GNU \texttt{gcc} v7.6.0\\ Apple \texttt{clang} v12.0.5\end{tabular} \\
\textbf{C++} & \textasciitilde & \checkmark & \checkmark & \begin{tabular}[c]{@{}l@{}}GNU \texttt{g++} v7.6.0\\ Apple \texttt{clang++} v12.0.5\end{tabular} \\
\rowcolor[HTML]{EFEFEF} 
\textbf{C\#} & \textasciitilde & \checkmark & \checkmark & .NET 5.0 via \texttt{dotnet-script} \\
\textbf{JavaScript} & \textasciitilde & \checkmark & \checkmark & Node.js v16.4.1 \\
\rowcolor[HTML]{EFEFEF} 
\textbf{Java} & \textasciitilde & \checkmark & \checkmark & OpenJDK v16.0.1 \\
\textbf{Rust} & \textasciitilde & \checkmark & \checkmark & \texttt{rustc} v1.53.0 \\
\rowcolor[HTML]{EFEFEF} 
\textbf{Go} & \textasciitilde & \checkmark & \checkmark & \texttt{go} v1.16.6 \\
\textbf{Python} & \checkmark & \checkmark & \checkmark & \begin{tabular}[c]{@{}l@{}}Python 3.9.5 via \texttt{clang}\\ Python 3.7.10 via \texttt{gcc}\end{tabular} \\
\rowcolor[HTML]{EFEFEF} 
\textbf{SQL} & \checkmark & \checkmark & \checkmark & SQLite v3.39.4 \\
\textbf{Bash} & \textasciitilde & \checkmark & \checkmark & zsh v5.8.1 \\
\rowcolor[HTML]{EFEFEF} 
\textbf{Assembly} & \checkmark & \checkmark & \textasciitilde & x86\_64 gas on Apple clang v14.0.0 \\
\textbf{Solidity} & \checkmark & \checkmark & \textasciitilde & Solidity v0.8.16 \\\bottomrule
\end{tabular}
\end{table*}

%% file: tables/viewer-evaluation.tex
\begin{table*}[t]
\centering
\caption{Evaluation of common code editors and web-based repository front-ends for Trojan-Source-vulnerable rendering.\\Vulnerable visualizations at the time of discovery are marked with \checkmark and software patched after disclosure is shaded.}
\label{tab:viewers}
\resizebox{\textwidth}{!}{%
\begin{tabular}{@{}llllllllllllll}
\toprule
 & Visual Studio Code & Atom & SublimeText & Notepad++ & Eclipse & IntelliJ & Visual Studio & Xcode & vim & emacs & GitHub & BitBucket & GitLab \\ \midrule
\textbf{Windows} &  &  &  &  &  &  &  &  &  &  &  &  &  \\
Bidi Attack & \cellcolor{gray!20}\checkmark & \checkmark & \cellcolor{gray!20}Bidi unactioned & Displays control symbol & Mangled & Displays control char & \cellcolor{gray!20}Mangled & N/A & Mangled & \checkmark & \cellcolor{gray!20}\begin{tabular}[c]{@{}l@{}}Chrome: \checkmark\\ Firefox: \checkmark\\ Edge: \checkmark\end{tabular} & \cellcolor{gray!20}\begin{tabular}[c]{@{}l@{}}Chrome: \checkmark\\ Firefox: \checkmark\\ Edge: \checkmark\end{tabular} & \cellcolor{gray!20}\begin{tabular}[c]{@{}l@{}}Chrome: \checkmark\\ Firefox: \checkmark\\ Edge: \checkmark\end{tabular} \\
Homoglyph Attack & \cellcolor{gray!20}\checkmark & \checkmark & \checkmark & \checkmark & Missing Glyph & \checkmark & \checkmark & N/A & Misrendered & \checkmark & \begin{tabular}[c]{@{}l@{}}Chrome: \checkmark\\ Firefox: \checkmark\\ Edge: \checkmark\end{tabular} & \begin{tabular}[c]{@{}l@{}}Chrome: \checkmark\\ Firefox: \checkmark\\ Edge: \checkmark\end{tabular} & \cellcolor{gray!20}\begin{tabular}[c]{@{}l@{}}Chrome: \checkmark\\ Firefox: \checkmark\\ Edge: \checkmark\end{tabular} \\ \midrule
\textbf{MacOS} &  &  &  &  &  &  &  &  &  &  &  &  &  \\
Bidi Attack & \cellcolor{gray!20}\checkmark & \checkmark & Bidi unactioned & N/A & \checkmark & Displays control char & \checkmark & \checkmark & Displays codepoint & Displays underscores & \cellcolor{gray!20}\begin{tabular}[c]{@{}l@{}}Chrome: \checkmark\\ Firefox: \checkmark\\ Edge: \checkmark\\ Safari: Wrong order\end{tabular} & \cellcolor{gray!20}\begin{tabular}[c]{@{}l@{}}Chrome: \checkmark\\ Firefox: \checkmark\\ Edge: \checkmark\\ Safari: Wrong order\end{tabular} & \cellcolor{gray!20}\begin{tabular}[c]{@{}l@{}}Chrome: \checkmark\\ Firefox: \checkmark\\ Edge: \checkmark\\ Safari: Wrong order\end{tabular} \\
Homoglyph Attack & \cellcolor{gray!20}\checkmark & \checkmark & \checkmark & N/A & \checkmark & \checkmark & \checkmark & \checkmark & \checkmark & \checkmark & \begin{tabular}[c]{@{}l@{}}Chrome: \checkmark\\ Firefox: \checkmark\\ Edge: \checkmark\\ Safari: \checkmark\end{tabular} & \begin{tabular}[c]{@{}l@{}}Chrome: \checkmark\\ Firefox: \checkmark\\ Edge: \checkmark\\ Safari: \checkmark\end{tabular} & \cellcolor{gray!20}\begin{tabular}[c]{@{}l@{}}Chrome: \checkmark\\ Firefox: \checkmark\\ Edge: \checkmark\\ Safari: \checkmark\end{tabular} \\ \midrule
\textbf{Ubuntu} &  &  &  &  &  &  &  &  &  &  &  &  &  \\
Bidi Attack & \cellcolor{gray!20}\checkmark & \checkmark & \cellcolor{gray!20}Bidi unactioned & N/A & \checkmark & Displays control char & N/A & N/A & Displays codepoint & \checkmark & \cellcolor{gray!20}\begin{tabular}[c]{@{}l@{}}Chrome: \checkmark\\ Firefox: \checkmark\end{tabular} & \cellcolor{gray!20}\begin{tabular}[c]{@{}l@{}}Chrome: \checkmark\\ Firefox: \checkmark\end{tabular} & \cellcolor{gray!20}\begin{tabular}[c]{@{}l@{}}Chrome: \checkmark\\ Firefox: \checkmark\end{tabular} \\
Homoglyph Attack & \cellcolor{gray!20}\checkmark & \checkmark & \checkmark & N/A & \checkmark & \checkmark & N/A & N/A & \checkmark & \checkmark & \begin{tabular}[c]{@{}l@{}}Chrome: \checkmark\\ Firefox: \checkmark\end{tabular} & \begin{tabular}[c]{@{}l@{}}Chrome: \checkmark\\ Firefox: \checkmark\end{tabular} & \cellcolor{gray!20}\begin{tabular}[c]{@{}l@{}}Chrome: \checkmark\\ Firefox: \checkmark\end{tabular} \\ \bottomrule
\end{tabular}%
}
\end{table*}

%% file: sections/discussion.tex
\section{Discussion}

\input{figures/rust-3}

\subsection{Ethics}

We followed our department's ethical guidelines carefully during this research. We did not launch any attacks using Trojan Source methods against codebases we did not own. Furthermore, we performed responsible disclosure to all companies and organizations owning products in which we discovered vulnerabilities. We negotiated a 99-day embargo period following our first disclosure to allow affected products to be repaired, and we will discuss that process later.

\subsection{Attack Feasibility}

Attacks on source code are very attractive and valuable to motivated adversaries, as maliciously inserted backdoors can be incorporated into signed code that persists in the wild for long periods of time. Moreover, if backdoors are inserted into open source software components that are included downstream by many other applications, the blast radius of such an attack can be very large. Trojan Source attacks introduce the possibility of inserting such vulnerabilities into source code invisibly, thus completely circumventing the current principal control against them, namely human source code review. 
There is a long history of the attempted insertion of backdoors into critical code bases. One example was the attempted insertion of a root user escalation-of-privilege backdoor into the Unix kernel, which was as subtle as changing an \texttt{==} token to an \texttt{=} token~\cite{lwnbackdoor}. This attack was detected when experienced developers saw the vulnerability. The techniques described here allow such attacks to be harder to detect in future.

Recent research in developer security usability has shown that a significant portion of developers will gladly copy and paste insecure source code from unofficial online sources such as Stack Overflow\footnote{\href{https://stackoverflow.com}{stackoverflow.com}}~\cite{7546508,7958574}. Since Bidi control characters persist through copy-and-paste functionality, malicious code snippets with invisible vulnerabilities can be posted online in the hope that they will end up in production code. The market for such vulnerabilities is vibrant, with exploits on major platforms now commanding seven-figure sums~\cite{Perlroth2021}.

As of the time of discovery, C, C++, C\#, JavaScript, Java, Rust, Go, Python, SQL, Bash, Assembly, and Solidity were all vulnerable to Trojan Source attacks. They are all still formally vulnerable at the time of writing as their specifications are unchanged, although some of their compilers or interpreters have now implemented defenses. More broadly, this class of attacks is likely applicable to any language with common compilers that accept Unicode source code. Any entity whose security relies on the integrity of software supply chains should be concerned. 

\subsection{Syntax Highlighting}

Many developers use text editors that, in addition to basic text editing features, provide syntax highlighting for the languages in which they are programming. Moreover, many code repository platforms, such as GitHub\footnote{\href{https://github.com}{github.com}}, provide syntax highlighting through a web browser. Comments are often displayed in a  different color from code, and many of the proofs of concept provided in this paper work by deceiving developers into thinking that comments are code or vice versa.

We might have hoped that a well-implemented syntax-highlighting platform would at the very least exhibit unusual syntax highlighting in the vicinity of Bidi control characters in code, but our experience at the time of discovery was mixed. Some attacks provided strange highlighting in a subset of editors, but all syntax highlighting nuances depended on both the editor and the attack.

Although unexpected coloring of source code may flag the possibility of an encoding attack to experienced developers, especially once they are familiar with this work, we expect that most developers would not even notice unusual highlighting, let alone investigate it thoroughly enough to work out what was going on. A motivated attacker could experiment with the visualization of different attacks in the text editors and code repository front-ends used in targeted organizations in order to select an attack that has no or minimal visual effect.

Bidi control characters will typically cause a cursor to jump positions on a line when using arrow keys to click through tokens, or to highlight a line of text character-by-character. This is an artifact of the effect of the logical ordering of tokens on many operating systems and Unicode implementations. Such behavior, while producing no visible changes in text, may also be enough to alert some experienced developers. However, we suspect that this requires more attention than is given by most developers to reviews of large pieces of code.

\subsection{Invisible Character Attacks}

When discussing the string-stretching technique, we proposed that invisible characters or homoglyphs could be used to make visually-identical strings that are logically different when compared. Another invisible-vulnerability technique with which we experimented -- largely without success -- was the use of invisible characters in function names.

We theorized that invisible characters included in a function name could define a different function from the function defined by only the visible characters. This could allow an attacker to define an adversarial version of a standard function, such as \texttt{printf} in C, that can be invoked by calling the function with an invisible character in the function name. Such an adversarial function definition could be discreetly added to a codebase by defining it in a common open-source package that is imported into the global namespace of the target program.

However, we found that all compilers analyzed in this paper emitted compilation errors when this technique was employed, with the exception of Apple \texttt{clang} v12.0.5 (which emitted a warning instead of an error), SQL, and shell scripts on zsh.

Should a compiler not instrument defenses against invisible characters in function definition names -- or indeed in variable names -- this attack may well be feasible. That said, our experimental evidence suggests that this theoretical attack already has defenses employed against it by most modern compilers, and thus is unlikely to work in practice.

Following the public disclosure of Trojan Source attacks, open source contributors suggested another invisible character attack. In this attack, an adversary uses an invisible character to divide multiline comment terminating sequences. By doing so, compilers typically won't close the comment and the subsequent lines are not executed thus creating a variant of the Commenting-Out technique. An example of this attack in Rust can be found in \Cref{fig:rust-3-encoded,fig:rust-3-rendered}. This example does not print any output because the entire function body is interpreted as a comment.

\subsection{Homoglyph Attacks}

After we investigated invisible characters, we wondered whether homoglyphs in function names could be used to define distinct functions whose names appeared to the human eye to be the same. Then an adversary could write a function whose name appears the same as a pre-existing function -- except that one letter is replaced with a visually similar character. Indeed, this same technique could be used on code identifiers of any kind, such as variables and class names, and may be particularly insidious for homoglyphs that appear like numbers. This attack likely falls under the heading of CWE 1007~\cite{mitre_cwe_1007}.

We were able to successfully implement homoglyph attack proofs-of-concept in every language discussed in this paper except Assembly and Solidity; that is, C, C++, C\#, JavaScript, Java, Rust, Go, Python, SQL, and Bash all appear to be vulnerable. In our experiments, we defined two functions that appeared to have the name \texttt{sayHello}, except that one version used a Latin H while the other used a Cyrillic \cyrillicH.

Consider \Cref{fig:c++-3}, which implements a homoglyph attack in C++. For clarity, we denote the Latin H in blue and the Cyrillic \cyrillicH\ in red. This program outputs the text \textit{Goodbye, World!} when compiled using \texttt{clang++}. Although this example program appears harmless, a homoglyph attack could cause significant damage when applied against a common function, perhaps via an imported library. For example, suppose a function called \texttt{hashPassword} was replaced with a similar function that called and returned the same value as the original function, but only after leaking the pre-hashed password over the network.

All compilers and interpreters examined in this paper emitted the text \textit{Goodbye, World!} with similar proofs of concept. There were only three exceptions. GNU's \texttt{gcc} and its C++ counterpart, \texttt{g++}, both emitted stray token errors. 
Of particular note is the Rust compiler, which threw a `mixed\_script\_confusables' warning while producing the homoglyph attack binary. The warning text suggested that the function name with the Cyrillic \cyrillicH\ used ``mixed script confusables'' and suggested rechecking to ensure usage of the function was wanted. This is a well-designed defense against homoglyph attacks, and it shows that this attack had been seriously considered by at least one compiler team. 

This defense, together with the defenses against invisible character attacks, should serve as a precedent. It is reasonable to expect compilers to also incorporate defenses against Trojan Source attacks. 

\input{figures/c++-3}

\subsection{Defenses}

The simplest defense is to ban the use of text directionality control characters both in language specifications and in compilers implementing these languages.

In most settings, this simple solution may well be sufficient. If an application wishes to print text that requires Bidi control characters, developers can generate those characters using escape sequences rather than embedding potentially dangerous characters into source code. 

This simple defense can be improved by adding a small amount of nuance. By banning all directionality-control characters, users with legitimate Bidi control character use cases in comments are penalized. Therefore, a better defense might be to ban the use of \textit{unterminated} Bidi control characters within string literals and comments. By ensuring that each control character is terminated -- that is, for example, that every \texttt{LRI} has a matching \texttt{PDI} -- it becomes impossible to distort legitimate source code outside of string literals and comments.

Trojan Source defenses must be enabled by default on all compilers that support Unicode input, and turning off the defenses should only be permitted when a dedicated suppression flag is passed.

While changes to language specifications and compilers are ideal solutions, there is an immediate need for existing code bases to be protected against this family of attacks. Moreover, some languages or compilers may choose not to implement appropriate defenses. To protect organizations that rely on them, defenses can be employed in build pipelines, code repositories, and text editors.

Build pipelines, such as those used by software producers to build and sign production code, can scan for the presence of Bidi control characters before initiating each build and break the build if such a character is found in source code. Alternatively, build pipelines can scan for the more nuanced set of unterminated Bidi control characters. Such tactics provide an immediate and robust defense for existing software maintainers.

Code repository systems and text editors can also help prevent Trojan Source attacks by making them visible to human reviewers. For example, code repository front-ends, such as web UIs for viewing committed code, can choose to represent Bidi control characters as visible tokens, thus making attacks visible, and by adding a visual warning to the affected lines of code.

Code editors can employ similar tactics. In fact, some already do; \texttt{vim}, for example, defaults to showing Bidi control characters as numerical code points rather than applying the Bidi algorithm. However, many common code editors did not adopt this behavior at the time of disclosure, including most GUI editors such as Microsoft's VS Code and Apple's Xcode.

Many of the largest compilers, code editors, and repositories adopted these defenses following a coordinated disclosure process; we will describe more detail, including caveats about false positives, later in this section.

\subsection{Compiler Responsibility}

The disclosure and release of Trojan Source attacks has sparked debate on whether compilers should protect against this vulnerability pattern.

Those advocating against argue that a compiler's job is to compile code, not to protect developers from all possible vulnerabilities. Linters, the argument follows, are the natural tool for exposing issues in code that deviate from standard form, and performing vulnerability checks here helps to keep compilers efficient.

Meanwhile, those advocating in favor argue that well-known vulnerabilities should be mitigated in compilers so that as much as possible of the ecosystem is inoculated against the attack. For example, most C compilers including GCC and clang emit warnings by default for any use of the unsafe \texttt{stdio} function \texttt{gets}; by the same logic, it is sensible to warn users of unsafe Bidi characters.

While Trojan Source attacks are strictly speaking a matter for the language rather than the compiler, we are of the view that compiler protections are in the best interest of the broader ecosystem.

\subsection{Ecosystem Scanning}

We were curious if we could find any examples of Trojan Source attacks in the wild prior to public disclosure of the attack vector, and therefore tried to scan as much of the open source ecosystem as we could for signs of attack.

We assembled a RegEx that identified unterminated Bidi control characters in comments and strings, and GitHub provided us with the results of this pattern run against all public commits containing non-markup language source code ingested into GitHub from January through mid October 2021 by internally running a Java-syntax RegEx\footnote{The exact RegEx used is available at \href{https://github.com/nickboucher/trojan-source/blob/main/RegEx/java.regex}{github.com/nickboucher/trojan-source/blob/main/RegEx/java.regex}. We provide a more readable RegEx in PCRE2 syntax as Appendix \Cref{fig:regex}.}against the relevant backend database. This yielded 7,444 commits of the over over 1 billion commits scanned, which resolved to 2,096 unique files still present in public repositories as of October 2021.

98.8\% of the results were false positives. Examples of clearly non-malicious encodings included LRE characters placed at the start of file paths, malformed strings in genuinely right-to-left languages, and Bidi characters placed into localized format string patterns. We note that these results do not imply that scanning for unterminated Bidi control characters as a compiler, code viewer, or repository defense is likely to yield a high false positive rate in practice. We also suspect that most of these false positives were generated by developer tooling that incorrectly injects Bidi characters to force a set text directionality. It is likely that such tools will be updated to use Unicode-compliant terminated Bidi sequences as Trojan Source defenses gain widespread adoption~\cite{unicode_bidi_2020}. Implementers of defenses should consider the conditions that cause false positives with this scanning technique and determine whether they are permissible in their setting.

However, we did find some evidence of techniques similar to Trojan Source attacks being exploited in 1.2\% of the GitHub RegEx scanning results. In one instance, a static code analysis tool for smart contracts, Slither~\cite{feist_slither_2018}, contained scanning for right-to-left override characters. The tool provides an example of why this scan is necessary: it uses an RLO character to swap the display order of two single-character variables passed as arguments. We also discovered multiple instances of JavaScript obfuscation that used Bidi characters to assist in obscuring code. This is not necessarily malicious, but is still an interesting use of directionality control characters. Frustratingly, we also discovered two instances of recipients of our embargoed disclosures experimenting with these attack techniques publicly prior to public release. Finally, our scans located multiple implementations of exploit generators for directionality control characters in filename extensions, as previously referenced~\cite{brian_krebs_right--left_2011}. Following public disclosure, we also discovered a GitHub issue referencing a technique similar to stretched-string attacks in the Go language repository, though the issue did not lead to a patch~\cite{github_issue_go}.

In parallel, contributors to the Rust project scanned all historical submissions to crates.io, Rust's package manager, and found no evidence of exploitation within the Rust ecosystem.

\subsection{Coordinated Disclosure}

We contacted nineteen independent companies and organizations in a coordinated disclosure effort to build defenses for affected compilers, interpreters, code editors, and code repository front-ends. We set a 99-day embargoed disclosure period during which disclosure recipients could implement defenses before we published our attacks. We met a variety of responses ranging from patching commitments and bug bounties to quick dismissal and references to legal policies.

We selected an initial set of disclosure recipients by identifying the maintainers of products that our experiments indicated were affected by the Trojan Source vulnerability pattern. We also included companies that, to our knowledge, maintained their own internal compilers and build tools. The initial disclosures were sent on July 25, 2021, and additional disclosures were sent as further impact was identified.


Of the nineteen software suppliers with whom we engaged, seven used an outsourced platform for receiving vulnerability disclosures, six had dedicated web portals for vulnerability disclosures, four accepted disclosures via PGP-encrypted email, and two accepted disclosures only via non-PGP email. They all confirmed receipt of our disclosure, and ultimately nine of them committed to releasing a patch.

Eleven of the recipients had bug bounty programs offering payment for vulnerability disclosures. Of these, five paid bounties, with an average payment of \$2,246.40 and a range of \$4,475.

On September 9, 2021, we sent a vulnerability report to CERT/CC, the CERT Coordination Center sponsored by CISA~\cite{certcc}. Our report was accepted the same day for coordinated disclosure assistance. This gave all affected vendors access to VINCE, a tool providing a shared communication platform across vendors implementing defenses. Thirteen of our recipients, including CERT/CC, opted in to the VINCE tool for these shared communications. CERT/CC also added three additional vendors to the disclosure beyond the nineteen we had already contacted.

On October 18, 2021, Trojan Source attacks were issued two CVEs~\cite{mitre_cve_2021}: CVE-2021-42574 for tracking the Bidi attack, and CVE-2021-42694 for tracking the homoglyph attack. These CVEs were issued by MITRE against the Unicode specification.

On the same day, we sent a PGP-encrypted disclosure to the \texttt{distros} mailing list~\cite{distros_list}, which contained representatives of the security teams of 21 operating systems at that time. This list coordinates the application of patches across OS maintainers with a maximum embargo period of 14 days.

We observed multiple patterns throughout the coordinated disclosure process:

\begin{itemize}
    \item \textbf{Novel Vulnerability Patterns} Vulnerability disclosures which do not follow commonly known vulnerability patterns (such as CWEs~\cite{mitre_cwe_2021}) are likely to be screened out by disclosure recipients. We observed a tendency to close issues immediately as representing no threat when they did not align to something well-known and easily demonstrated, such as SQL injection. 
    
    \item \textbf{Impactful Language} When writing vulnerability disclosures, descriptions that personalize the potential impact can be needed to drive action. Disclosures naming specific products at risk were most effective.
    
    \item \textbf{CVEs} CVEs are useful, as they increase the chance that the recipient will take the time to  read and understand the report. However, CVEs are by default raised by the affected supplier, so are not much help during an initial contact. We eventually had to fall back on the CVE issuer of last resort, MITRE.
    
    \item \textbf{Shared Communication} CERT/CC's VINCE platform provides a useful and neutral cross-organization discussion tool during coordinated disclosures. The tool allows affected vendors to post on a private discussion board, and makes it much easier to communicate to all affected parties in a central location. The CERT/CC team will also help to coordinate contacting affected vendors under embargo, which provides a helpful method for scaling out disclosure efforts at no cost. Like CVEs, having a CERT/CC case also adds to the credibility of disclosures.
    
    \item \textbf{Open Source Assistance} Disclosing to open source operating system security teams is helpful for assistance coordinating patches across the ecosystem, including with contributors of open source projects that may not otherwise offer an embargoed disclosure method. In particular, Linux operating systems backed by a commercial entity have both the funding and incentives to ensure that common open source tools are patched prior to public disclosure. Maintainers of open source projects commonly work for or closely with these companies, and as such can be included in security responses.
\end{itemize}

\subsection{Industry Response}

Multiple code repositories were patched following our disclosures. GitHub now shows a warning banner and code point visualization for Bidi control characters~\cite{github_advisory}. Likewise, Bitbucket now visualizes Bidi control characters as code points~\cite{atlassian_advisory}. GitLab, too, now visualizes both Bidi control characters and potential homoglyphs~\cite{gitlab_advisory}.

Likewise, some code editors also released patches. Visual Studio Code now visualizes both Bidi control characters and potential homoglyphs by default~\cite{vscode_advisory}. Emacs now uses a heuristic to apply visual highlighting to suspected malicious use of Bidi control characters~\cite{emacs_patch}.

Compiler patches were also published in response to the disclosures. Rust released a new set of lints in the \texttt{rustc} compiler that will throw errors upon detection of non-escaped Bidi characters~\cite{rust_advisory}. GCC released a new warning detecting Bidi misuse, \texttt{-Wbidi-chars}, which is enabled by default~\cite{gcc_wbidichars}. LLVM added a set of checks to the official \texttt{clang-tidy} linter, but did not add compiler warnings for performance considerations~\cite{llvm_advisory}. The Java team, on the other hand, declared the Trojan source attack to be an issue for code editors rather than for the language.

Finally, a new working group was proposed within Unicode to address Trojan Source-style attacks in future versions of the Unicode specification~\cite{unicode_avoiding_spoofing}.

%% file: figures/rust-3.tex
\begin{figure*}[t]
  \centering
  \begin{minipage}[b]{0.49\textwidth}
    \centering
    \frame{\includegraphics[width=.8\linewidth]{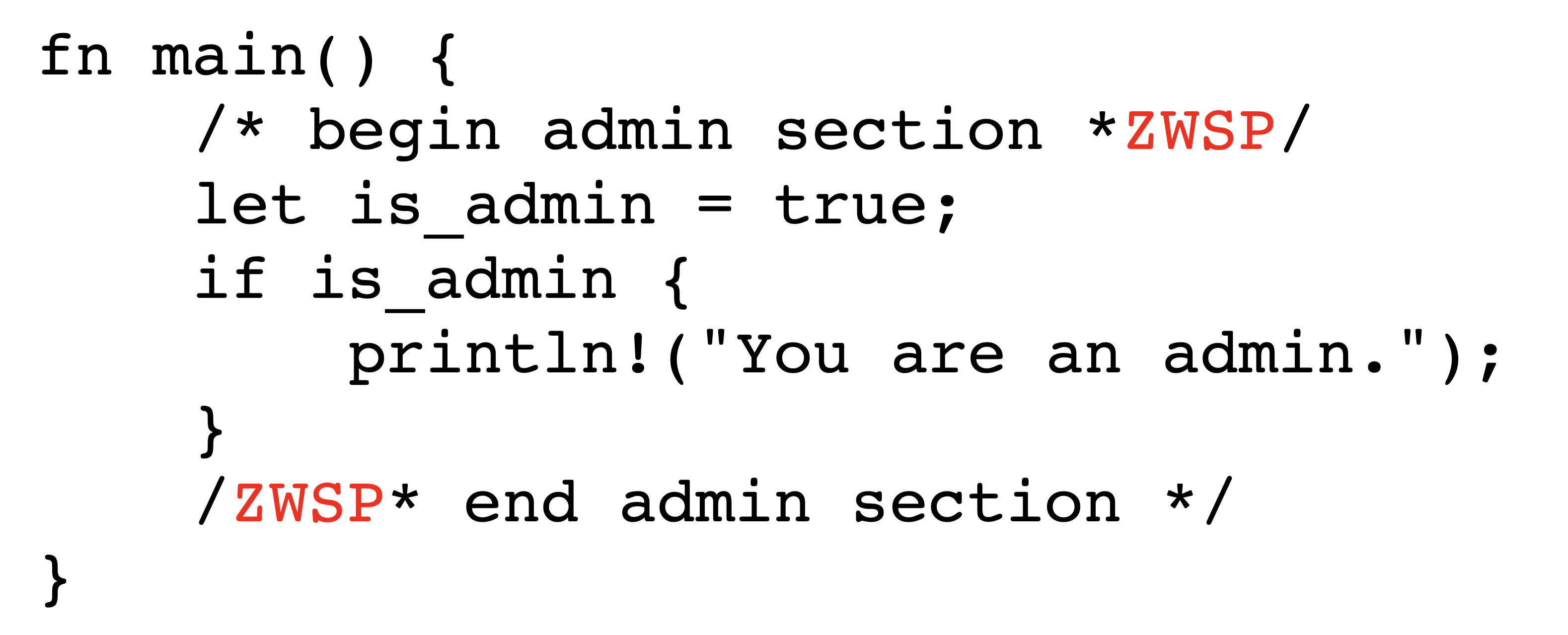}}
    \caption{Encoded bytes of a Trojan Source invisible character commenting-out attack in Rust.}
    \label{fig:rust-3-encoded}
  \end{minipage}
  \hfill
  \begin{minipage}[b]{0.49\textwidth}
    \centering
    \frame{\includegraphics[width=.8\linewidth]{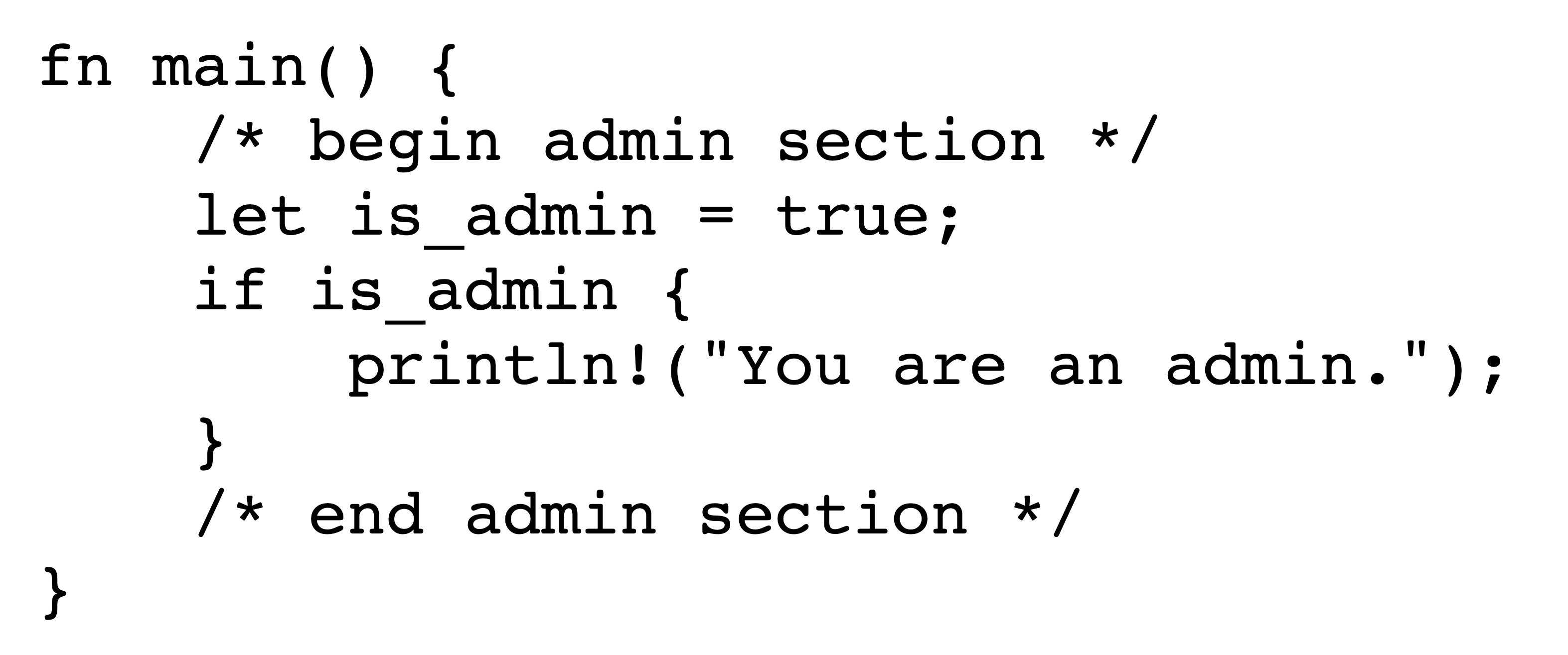}}
    \caption{Rendered text of a Trojan Source invisible character commenting-out attack in Rust.}
    \label{fig:rust-3-rendered}
  \end{minipage}
\end{figure*}

%% file: figures/c++-3.tex
\begin{figure}[t]
    \centering
    \frame{\includegraphics[width=.85\linewidth]{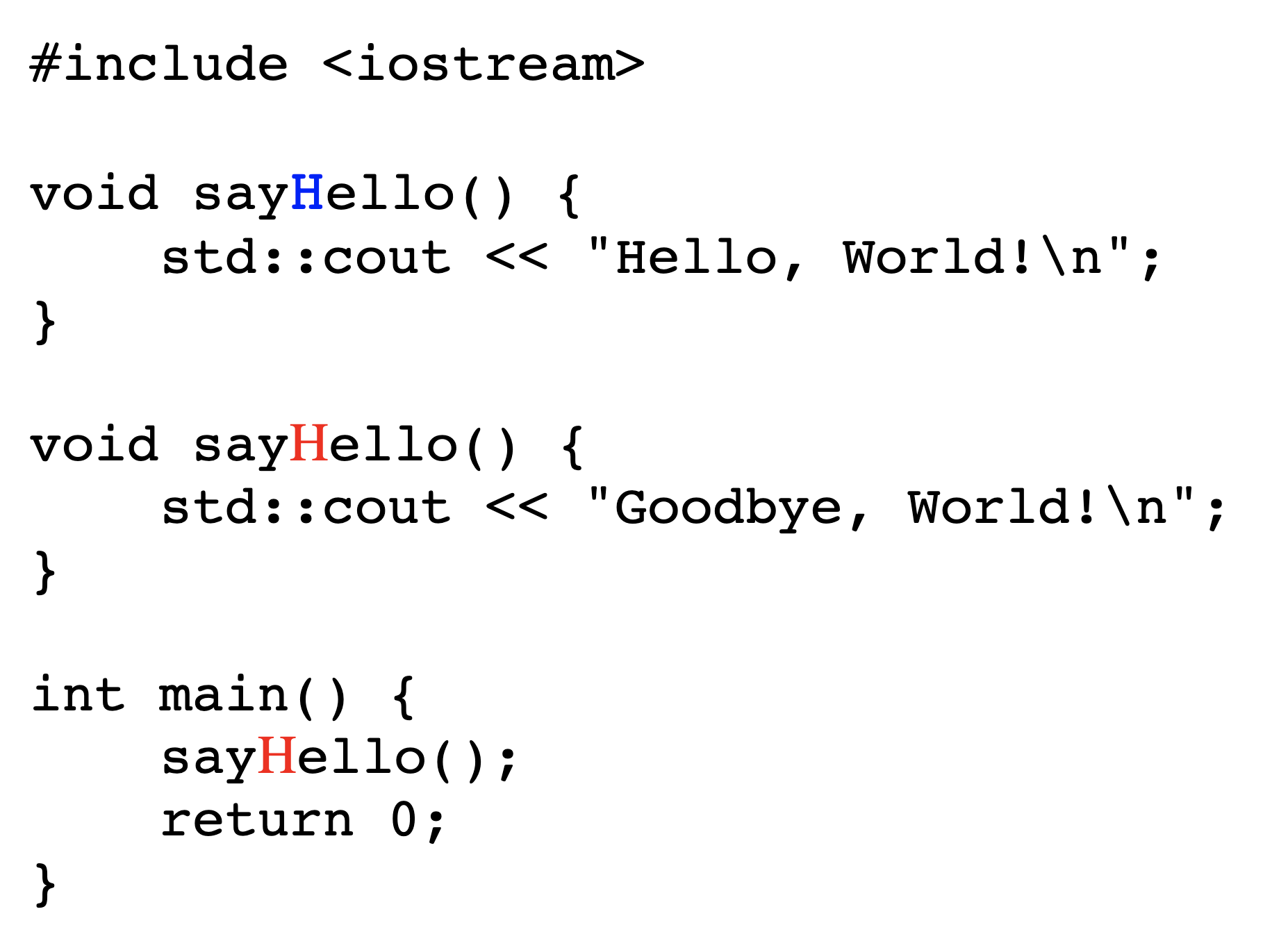}}
    \caption{Homoglyph function attack in C++.}
    \label{fig:c++-3}
\end{figure}

%% file: sections/conclusion.tex
\section{Conclusion}

We have presented a new type of attack that enables invisible vulnerabilities to be inserted into source code. Our Trojan Source attacks use Unicode control characters to modify the order in which blocks of characters are displayed, thus enabling comments and strings to appear to be code and vice versa. This enables an attacker to craft code that is interpreted one way by compilers and a different way by human reviewers. We present proofs of concept for C, C++, C\#, JavaScript, Java, Rust, Go, Python, SQL, Bash, and Assembly and argue that this attack may well appear in any programming language that supports internationalized text in comments and string literals, even in other encoding standards.

As powerful supply-chain attacks can be launched easily using these techniques, it is essential for organizations that participate in a software supply chain to implement defenses. We have discussed countermeasures that can be used at a variety of levels in the software development toolchain: the language specification, the compiler, the text editor, the code repository, and the build pipeline. We are of the view that the long-term solution to the problem will be deployed in compilers. We note that almost all compilers already defend against one related attack, which involves creating adversarial function names using zero-width space characters, while three generate errors in response to another, which exploits homoglyphs in function names. 

About half of the compiler maintainers we contacted during the disclosure period are working on patches or have committed to do so. As the others are dragging their feet, it is prudent to deploy other controls in the meantime where this is quick and cheap, or relevant and needful. Three firms that maintain code repositories are also deploying defenses. We recommend that governments and firms that rely on critical software should identify their their suppliers' posture, exert pressure on them to implement adequate defenses, and ensure that any gaps are covered by controls elsewhere in their toolchain. 

The fact that the Trojan Source vulnerability affects almost all computer languages makes it a rare opportunity for a system-wide and ecologically valid cross-platform and cross-vendor comparison of responses. As far as we are aware, it is an unprecedented test of the coordinated disclosure ecosystem.

Scientifically, this research also contributes to the growing body of work on security usability from the developer's perspective. It is not sufficient for a compiler to be verified; it must also be safely usable. Compilers that are trivially vulnerable to adversarial text encoding cannot reasonably be described as secure. 

%% file: sections/appendix.tex
\section{Appendix}

\vspace{1cm}

\subsection{C Trojan Source Proofs-of-Concept}
\label{apx:c}

\input{figures/c-1}
\input{figures/c-3}

\subsection{C++ Trojan Source Proofs-of-Concept}
\label{apx:c++}

\input{figures/c++-1}
\input{figures/c++-2}

\clearpage

\subsection{Java Trojan Source Proofs-of-Concept}
\label{apx:java}

\input{figures/java-1}
\input{figures/java-2}

\subsection{Rust Trojan Source Proofs-of-Concept}
\label{apx:rust}

\input{figures/rust-1}
\input{figures/rust-2}

\subsection{Python Trojan Source Proof-of-Concept}
\label{apx:py}

\input{figures/py-1}

\clearpage

\subsection{Go Trojan Source Proof-of-Concept}
\label{apx:go}

\input{figures/go-1}

\subsection{JavaScript Trojan Source Proof-of-Concept}
\label{apx:js}

\input{figures/js-2}

\subsection{Trojan Source Regular Expression}
\input{figures/regex}

%% file: figures/c-1.tex
\begin{figure*}[h]
  \centering
  \begin{minipage}[b]{0.49\textwidth}
    \centering
    \includegraphics[width=\linewidth]{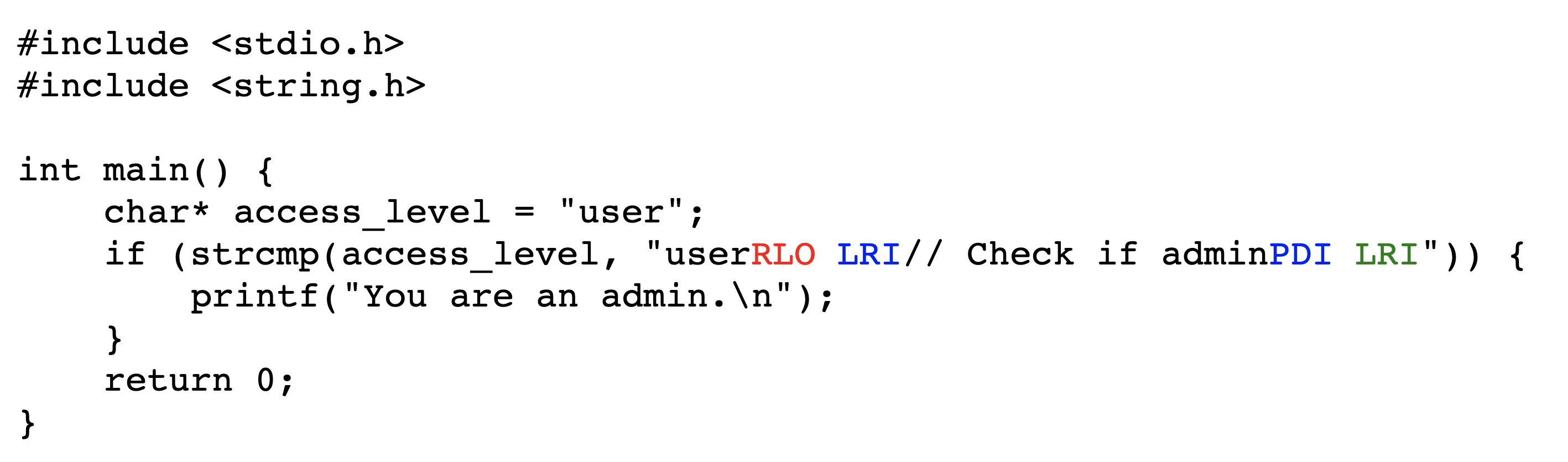}
    \caption{Encoded bytes of a Trojan Source stretched-string attack in C.}
    \label{fig:c-1-encoded}
  \end{minipage}
  \hfill
  \begin{minipage}[b]{0.49\textwidth}
    \centering
    \includegraphics[width=\linewidth]{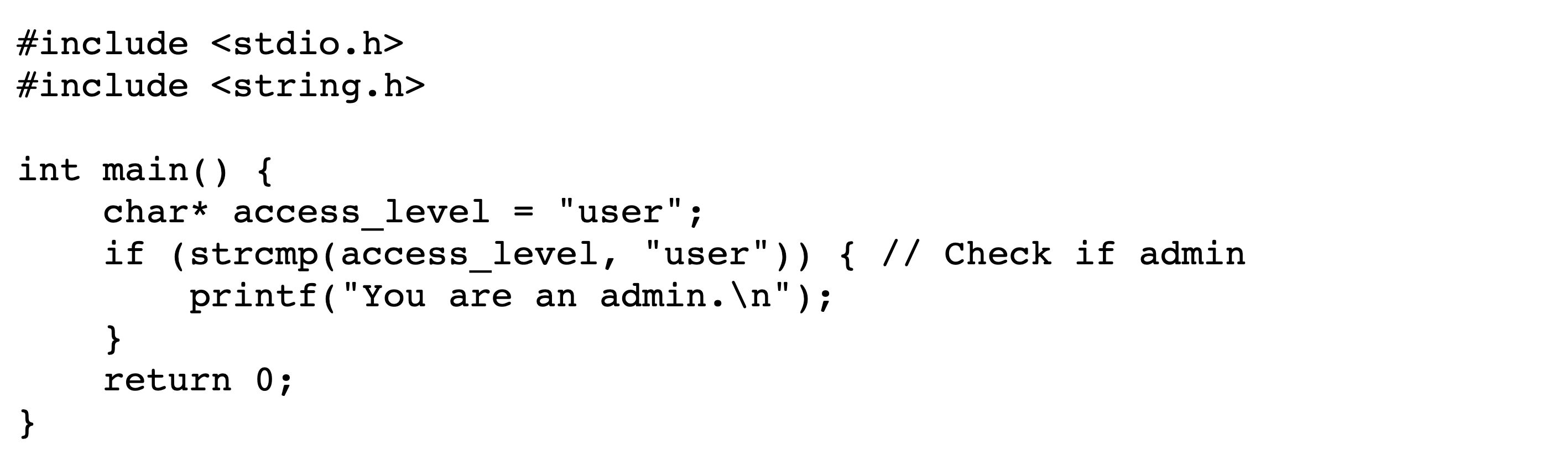}
    \caption{Rendered text of a Trojan Source stretched-string attack in C.}
    \label{fig:c-1-rendered}
  \end{minipage}
\end{figure*}

%% file: figures/c-3.tex
\begin{figure*}[h]
  \centering
  \begin{minipage}[b]{0.49\textwidth}
    \centering
    \includegraphics[width=\linewidth]{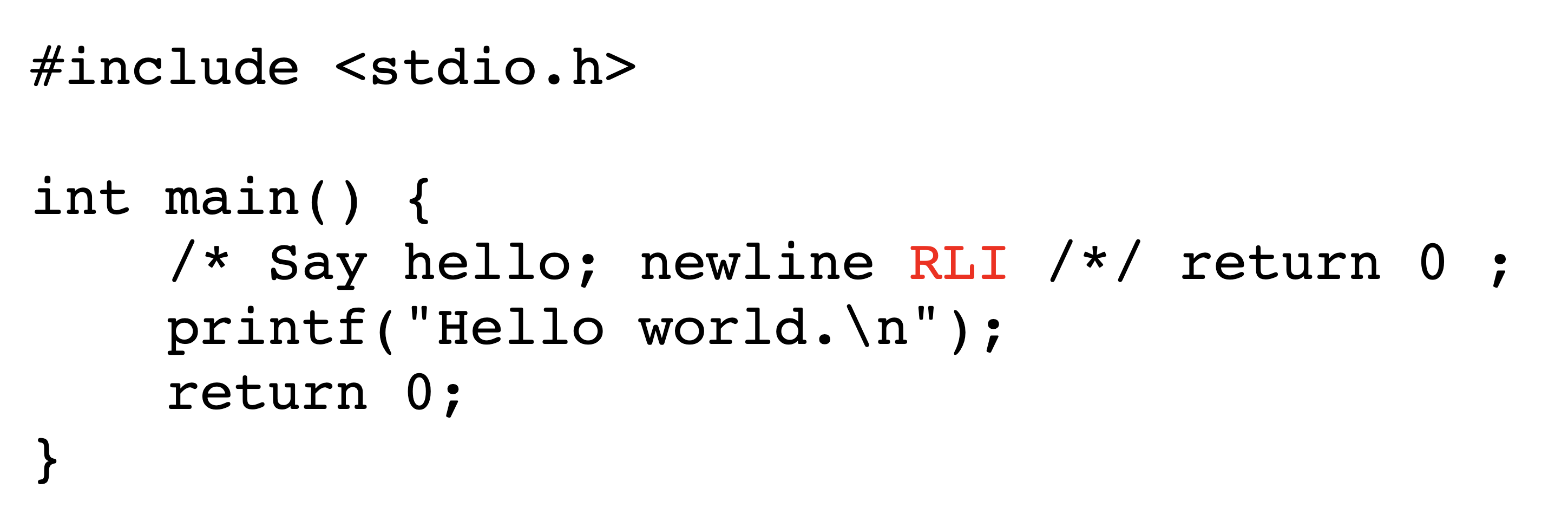}
    \caption{Encoded bytes of a Trojan Source early-return attack in C.}
    \label{fig:c-3-encoded}
  \end{minipage}
  \hfill
  \begin{minipage}[b]{0.49\textwidth}
    \centering
    \includegraphics[width=\linewidth]{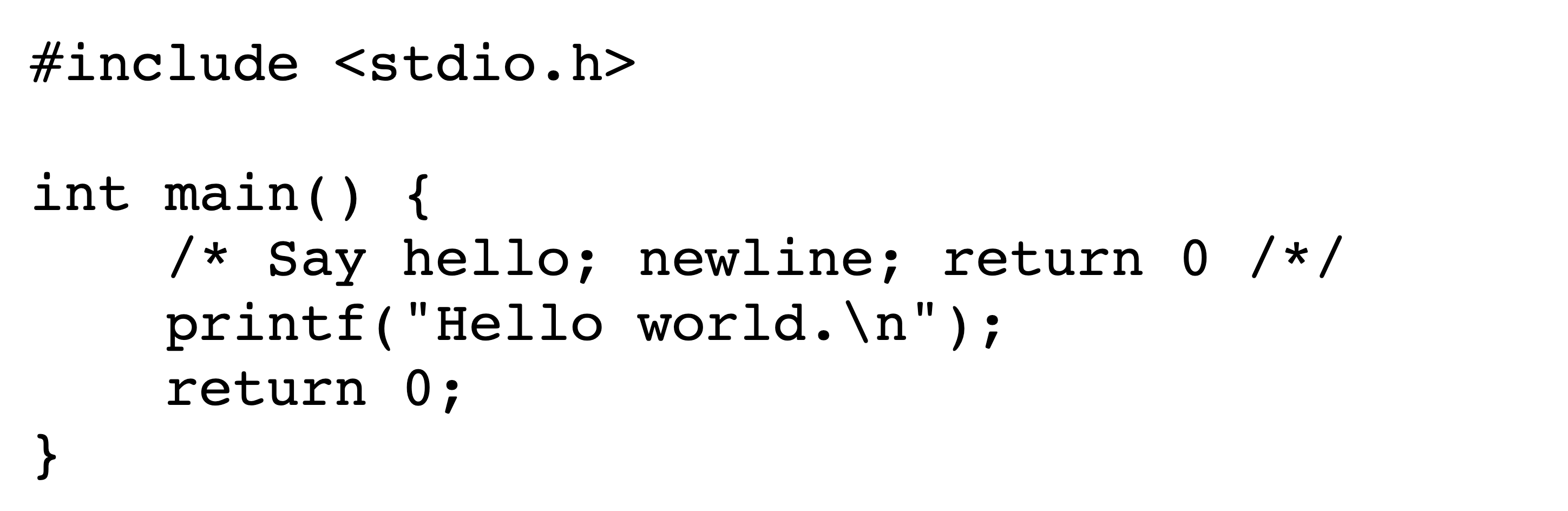}
    \caption{Rendered text of a Trojan Source early-return attack in C.}
    \label{fig:c-3-rendered}
  \end{minipage}
\end{figure*}

%% file: figures/c++-1.tex
\begin{figure*}[h]
  \centering
  \begin{minipage}[b]{0.49\textwidth}
    \centering
    \includegraphics[width=\linewidth]{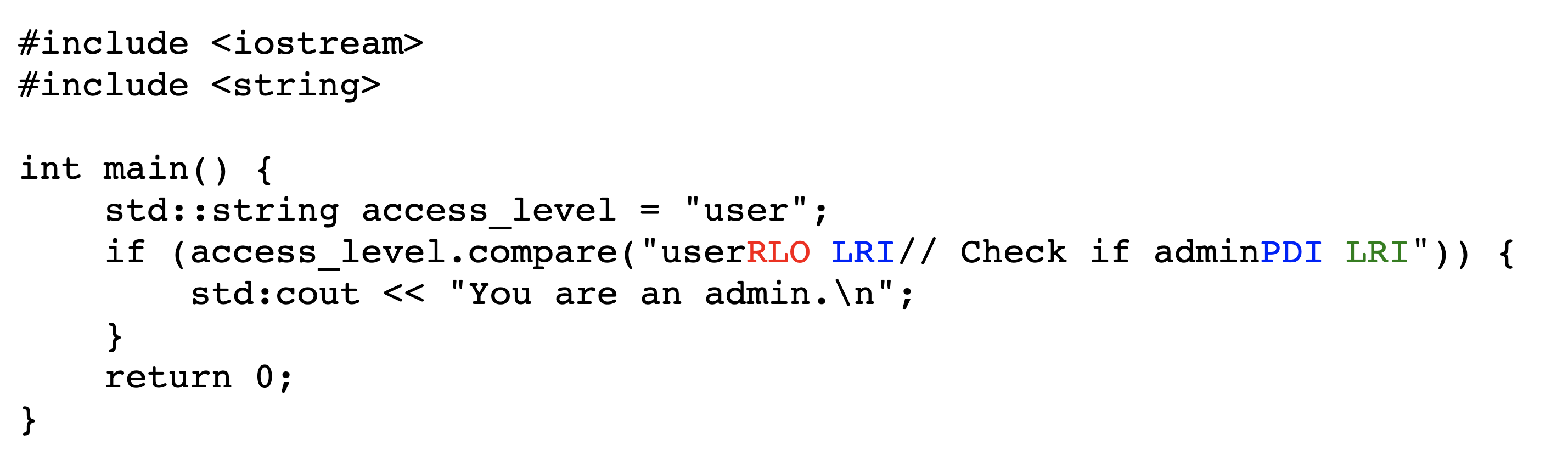}
    \caption{Encoded bytes of a Trojan Source stretched-string attack in C++.}
    \label{fig:c++-1-encoded}
  \end{minipage}
  \hfill
  \begin{minipage}[b]{0.49\textwidth}
    \centering
    \includegraphics[width=\linewidth]{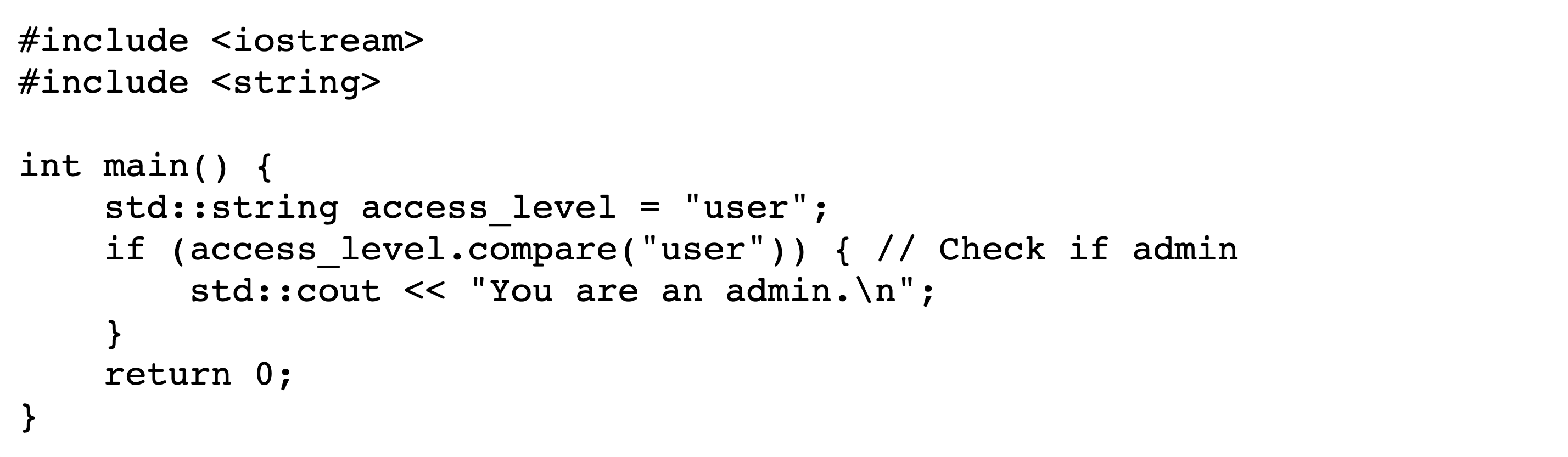}
    \caption{Rendered text of a Trojan Source stretched-string attack in C++.}
    \label{fig:c++-1-rendered}
  \end{minipage}
\end{figure*}

%% file: figures/c++-2.tex
\begin{figure*}[h!]
  \centering
  \begin{minipage}[b]{0.49\textwidth}
    \centering
    \includegraphics[width=\linewidth]{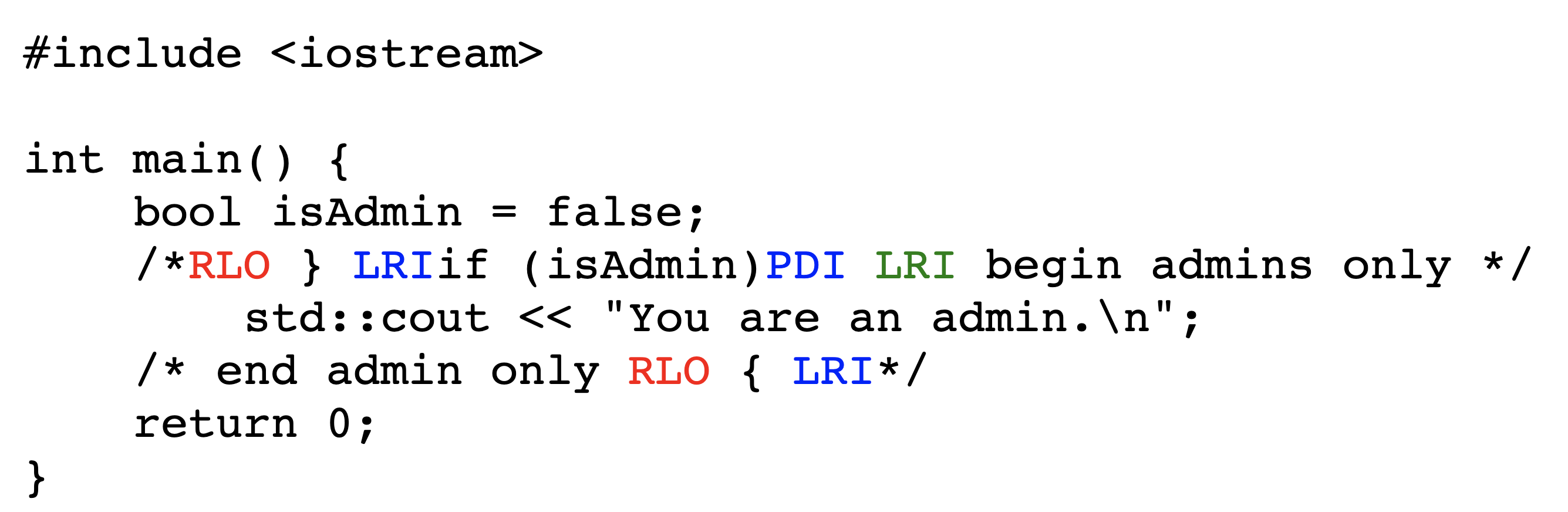}
    \caption{Encoded bytes of a Trojan Source commenting-out attack in C++.}
    \label{fig:c++-2-encoded}
  \end{minipage}
  \hfill
  \begin{minipage}[b]{0.49\textwidth}
    \centering
    \includegraphics[width=\linewidth]{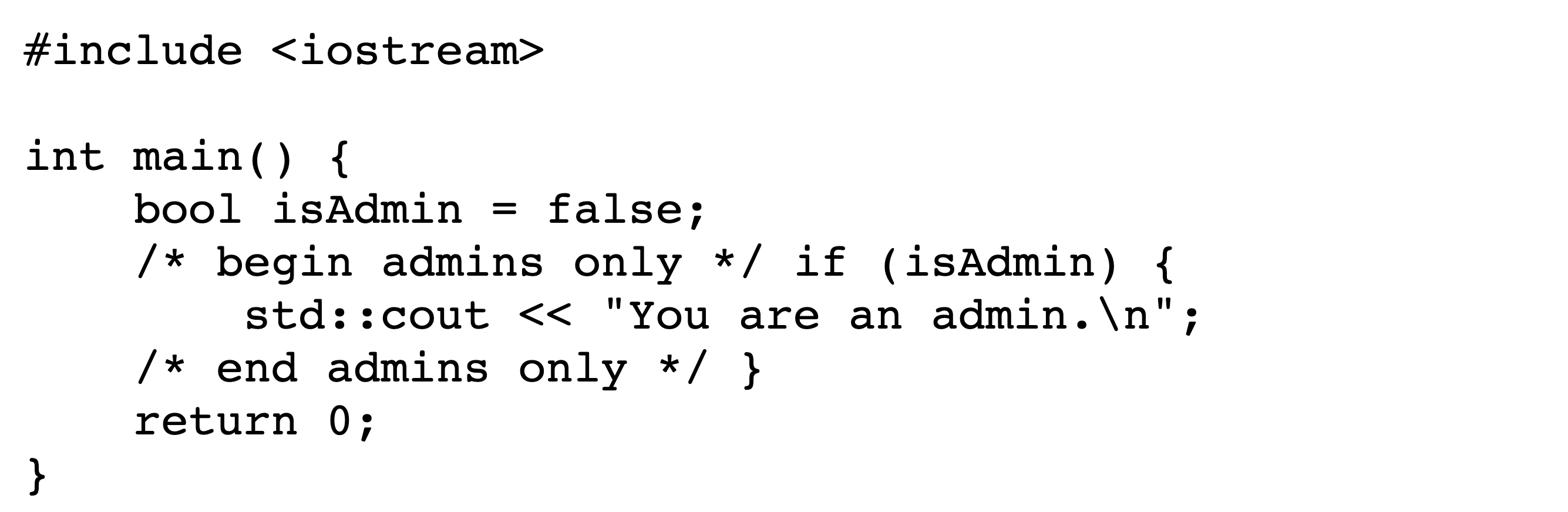}
    \caption{Rendered text of a Trojan Source commenting-out attack in C++.}
    \label{fig:c++-2-rendered}
  \end{minipage}
\end{figure*}

%% file: figures/java-1.tex
\begin{figure*}[h]
  \centering
  \begin{minipage}[b]{0.49\textwidth}
    \centering
    \includegraphics[width=\linewidth]{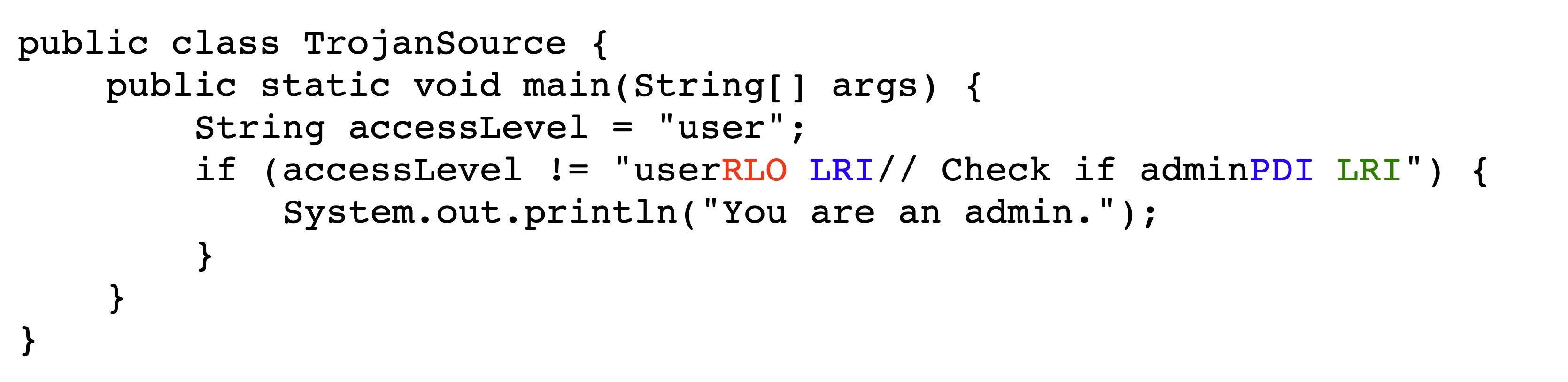}
    \caption{Encoded bytes of a Trojan Source stretched-string attack in Java.}
    \label{fig:java-1-encoded}
  \end{minipage}
  \hfill
  \begin{minipage}[b]{0.49\textwidth}
    \centering
    \includegraphics[width=\linewidth]{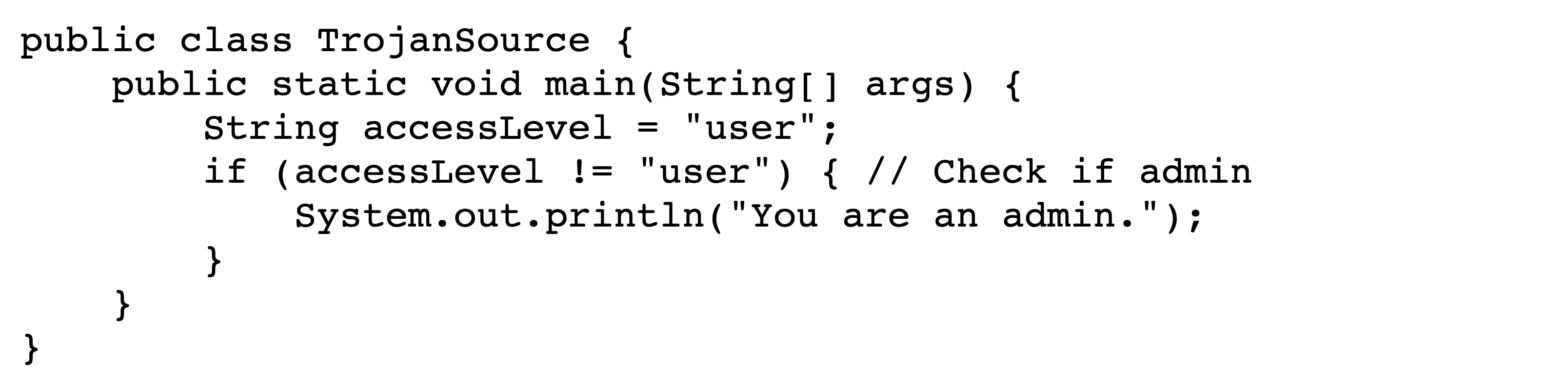}
    \caption{Rendered text of a Trojan Source stretched-string attack in Java.}
    \label{fig:java-1-rendered}
  \end{minipage}
\end{figure*}

%% file: figures/java-2.tex
\begin{figure*}[h]
  \centering
  \begin{minipage}[b]{0.49\textwidth}
    \centering
    \includegraphics[width=\linewidth]{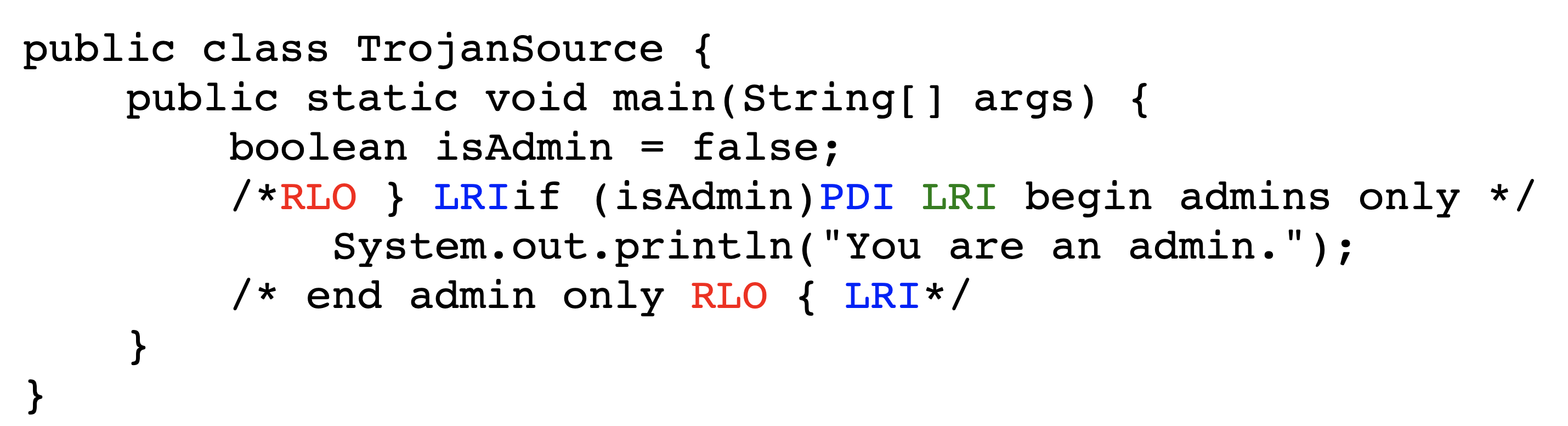}
    \caption{Encoded bytes of a Trojan Source commenting-out attack in Java.}
    \label{fig:java-2-encoded}
  \end{minipage}
  \hfill
  \begin{minipage}[b]{0.49\textwidth}
    \centering
    \includegraphics[width=\linewidth]{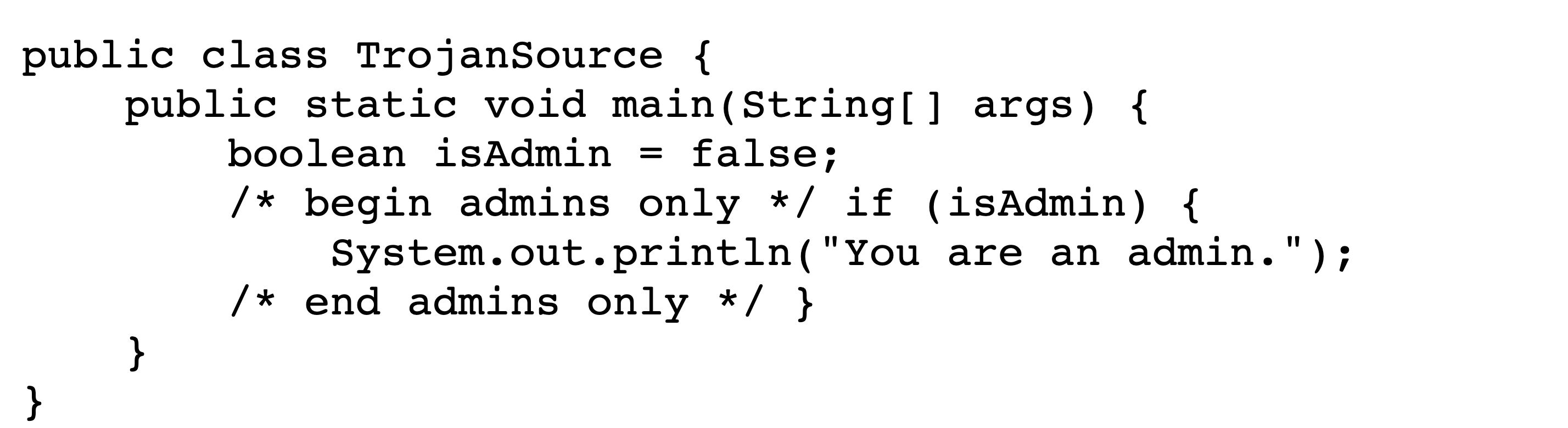}
    \caption{Rendered text of a Trojan Source commenting-out attack in Java.}
    \label{fig:java-2-rendered}
  \end{minipage}
\end{figure*}

%% file: figures/rust-1.tex
\begin{figure*}[h]
  \centering
  \begin{minipage}[b]{0.49\textwidth}
    \centering
    \includegraphics[width=\linewidth]{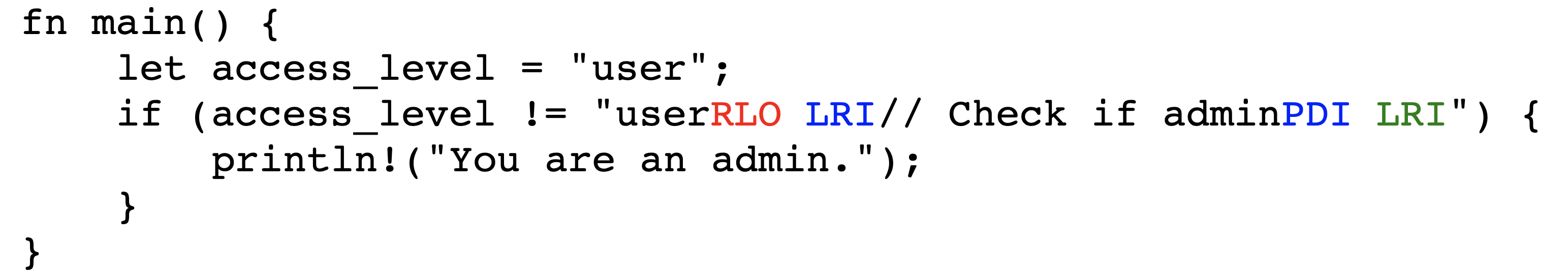}
    \caption{Encoded bytes of a Trojan Source stretched-string attack in Rust.}
    \label{fig:rust-1-encoded}
  \end{minipage}
  \hfill
  \begin{minipage}[b]{0.49\textwidth}
    \centering
    \includegraphics[width=\linewidth]{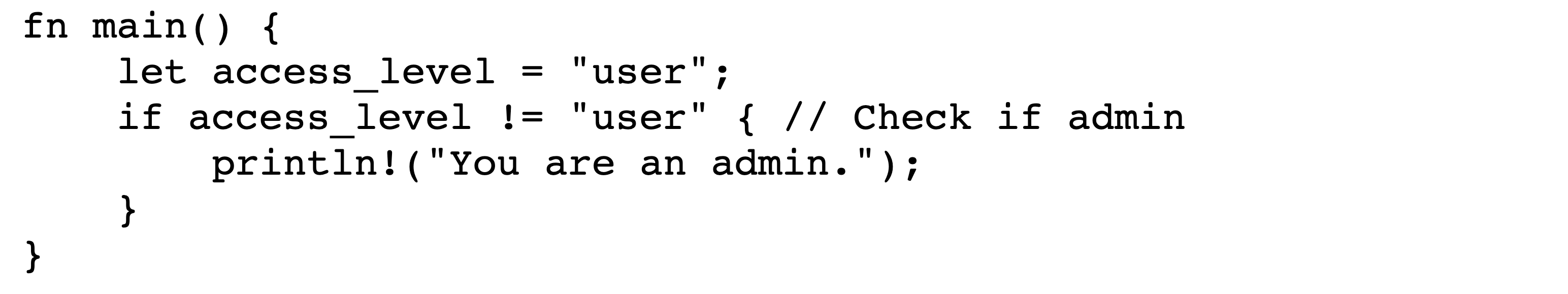}
    \caption{Rendered text of a Trojan Source stretched-string attack in Rust.}
    \label{fig:rust-1-rendered}
  \end{minipage}
\end{figure*}

%% file: figures/rust-2.tex
\begin{figure*}[h!]
  \centering
  \begin{minipage}[b]{0.49\textwidth}
    \centering
    \includegraphics[width=\linewidth]{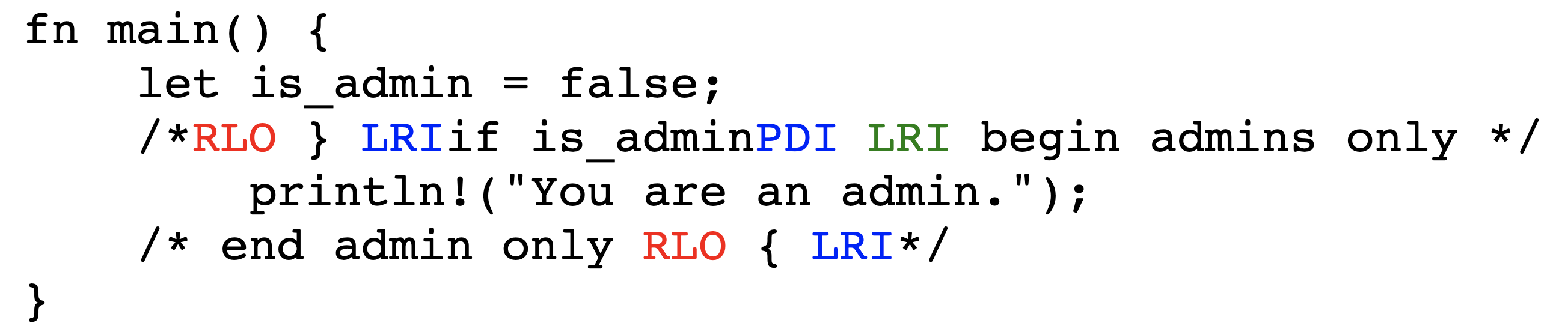}
    \caption{Encoded bytes of a Trojan Source commenting-out attack in Rust.}
    \label{fig:rust-2-encoded}
  \end{minipage}
  \hfill
  \begin{minipage}[b]{0.49\textwidth}
    \centering
    \includegraphics[width=\linewidth]{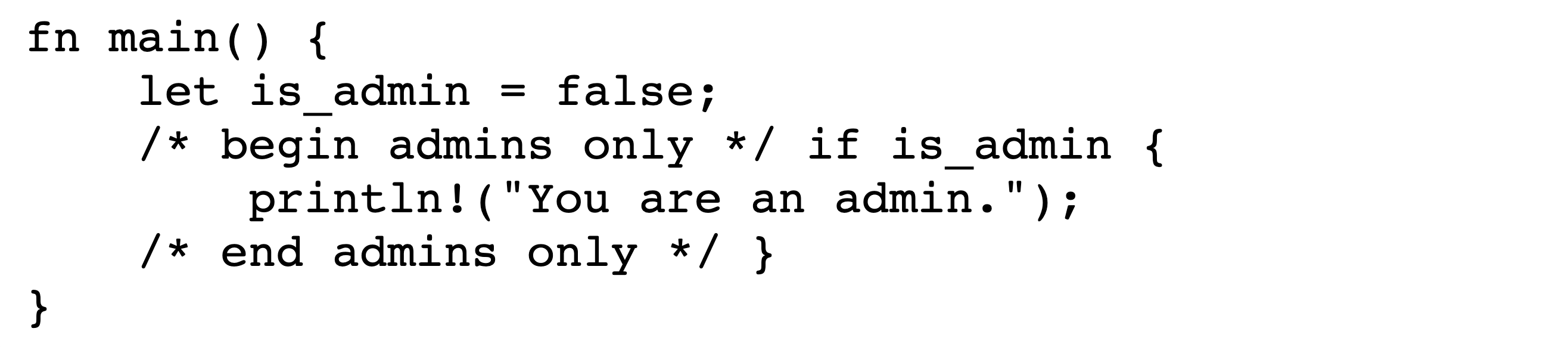}
    \caption{Rendered text of a Trojan Source commenting-out attack in Rust.}
    \label{fig:rust-2-rendered}
  \end{minipage}
\end{figure*}

%% file: figures/py-1.tex
\begin{figure*}[h!]
  \centering
  \begin{minipage}[b]{0.49\textwidth}
    \centering
    \includegraphics[width=\linewidth]{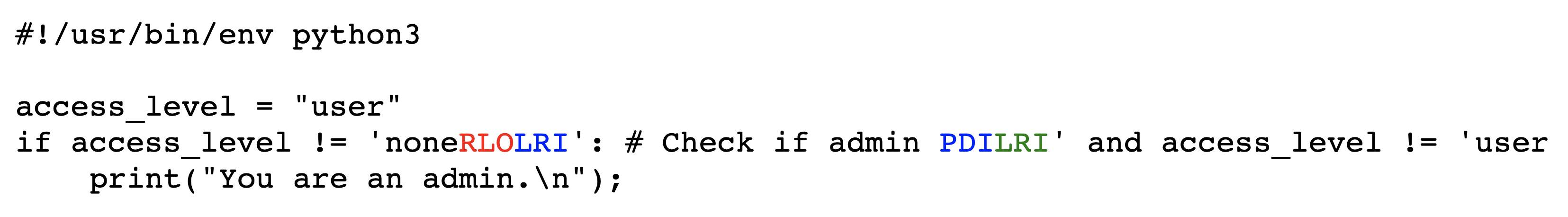}
    \caption{Encoded bytes of a Trojan Source commenting-out attack in Python.}
    \label{fig:py-1-encoded}
  \end{minipage}
  \hfill
  \begin{minipage}[b]{0.49\textwidth}
    \centering
    \includegraphics[width=\linewidth]{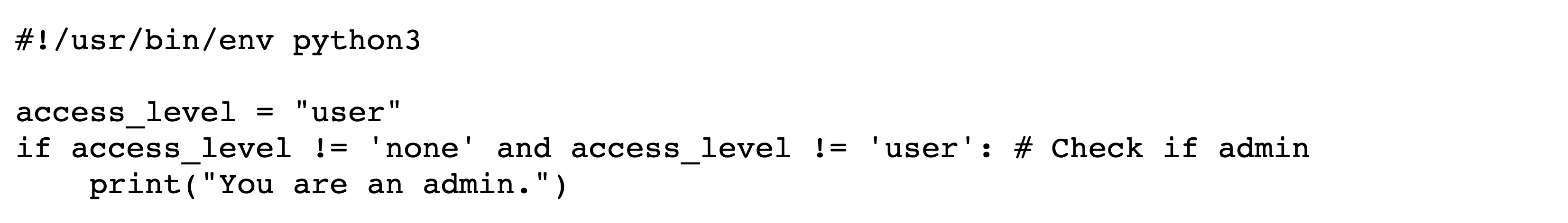}
    \caption{Rendered text of a Trojan Source commenting-out attack in Python.}
    \label{fig:py-1-rendered}
  \end{minipage}
\end{figure*}

%% file: figures/go-1.tex
\begin{figure*}[h!]
  \centering
  \begin{minipage}[b]{0.49\textwidth}
    \centering
    \includegraphics[width=\linewidth]{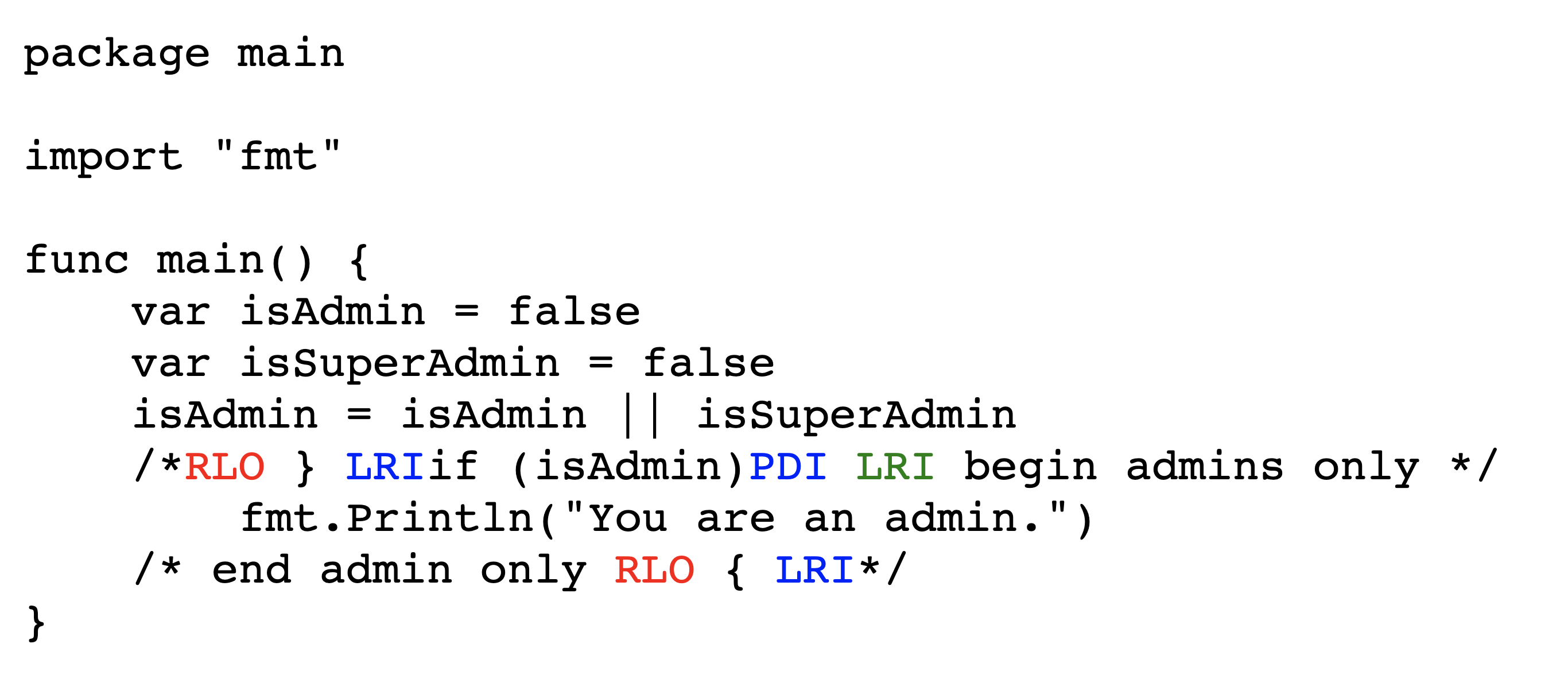}
    \caption{Encoded bytes of a Trojan Source commenting-out attack in Go.}
    \label{fig:go-1-encoded}
  \end{minipage}
  \hfill
  \begin{minipage}[b]{0.49\textwidth}
    \centering
    \includegraphics[width=\linewidth]{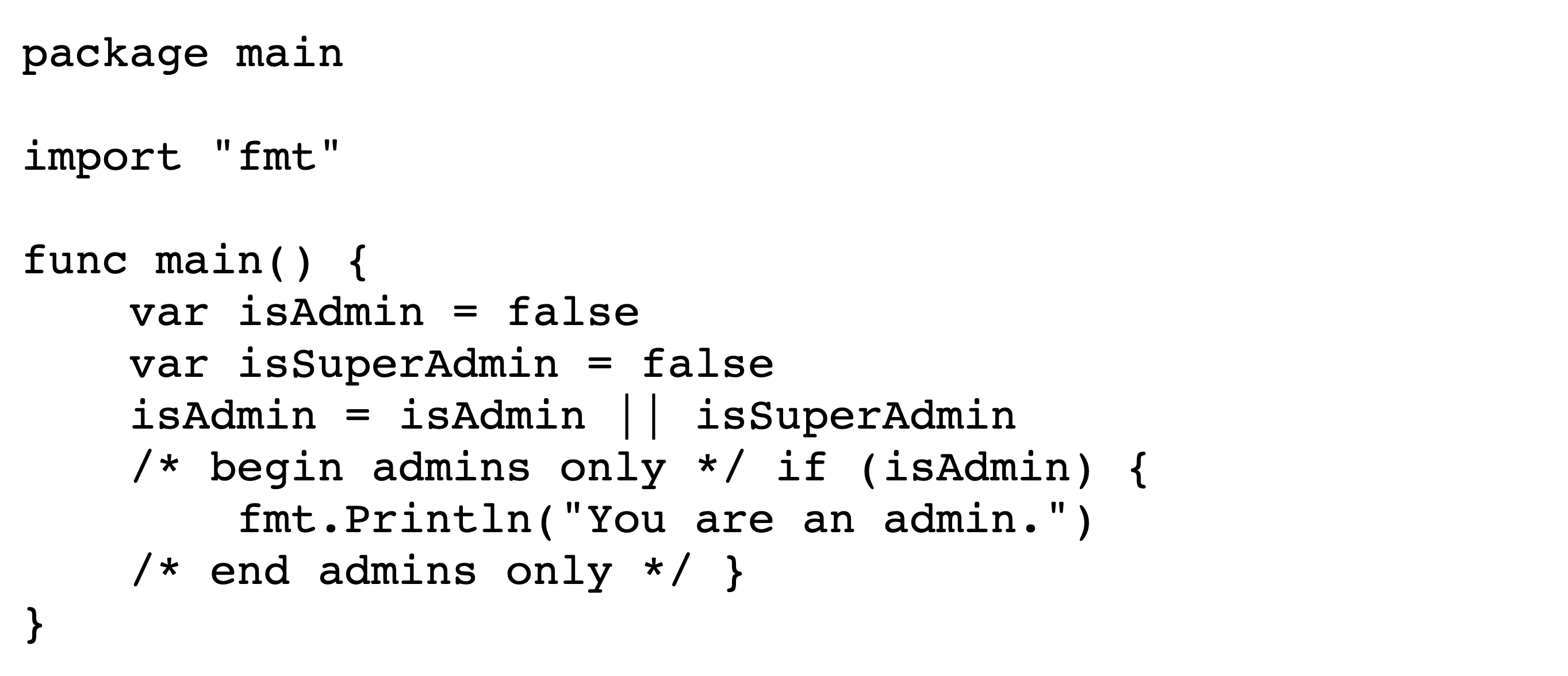}
    \caption{Rendered text of a Trojan Source commenting-out attack in Go.}
    \label{fig:go-1-rendered}
  \end{minipage}
\end{figure*}

%% file: figures/js-2.tex
\begin{figure*}[h]
  \centering
  \begin{minipage}[b]{0.49\textwidth}
    \centering
    \includegraphics[width=\linewidth]{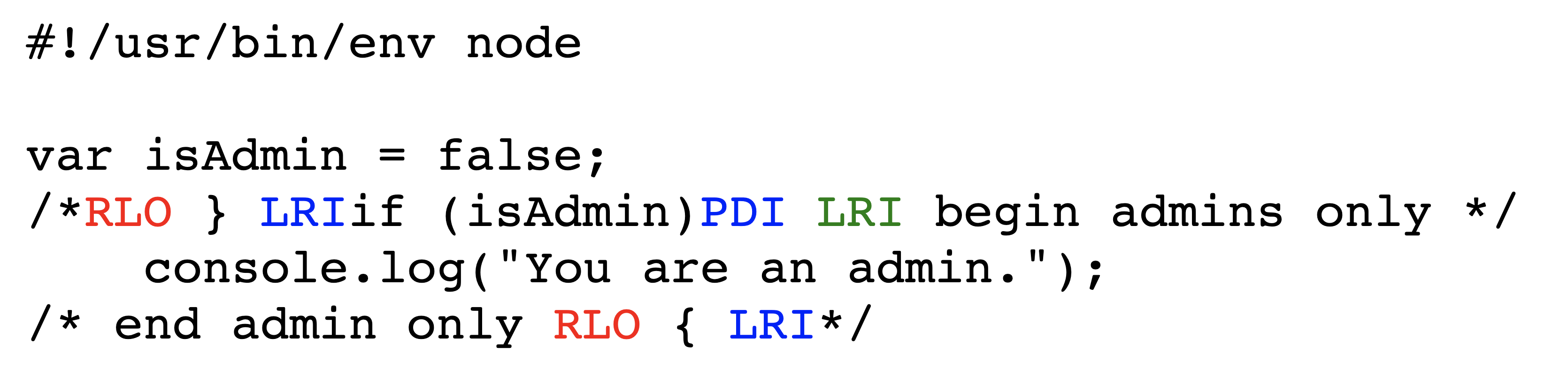}
    \caption{Encoded bytes of a Trojan Source commenting-out attack in JS.}
    \label{fig:js-2-encoded}
  \end{minipage}
  \hfill
  \begin{minipage}[b]{0.49\textwidth}
    \centering
    \includegraphics[width=\linewidth]{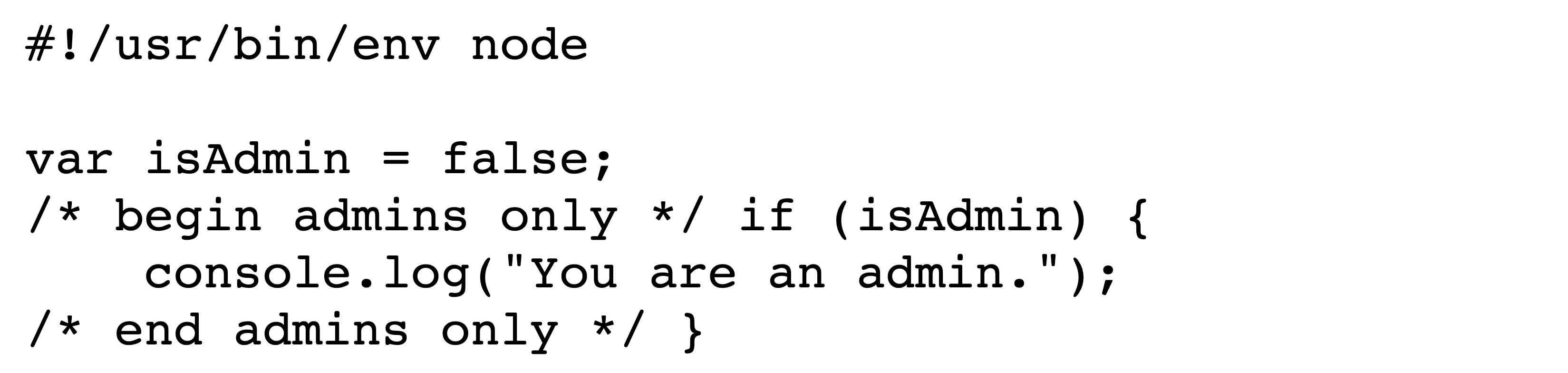}
    \caption{Rendered text of a Trojan Source commenting-out attack in JS.}
    \label{fig:js-2-rendered}
  \end{minipage}
\end{figure*}

%% file: figures/regex.tex
\begin{figure*}[h!]
\begin{lstlisting}
(?(DEFINE)
    (?<pdi>([^\x{2067}\x{2066}\x{2068}]*)([^\x{2067}\x{2066}\x{2068}\x{2069}]*)
            ((?-2)[\x{2067}\x{2066}\x{2068}](?-2)(?-1)*(?-2)[\x{2069}](?-2))*
            (?-3)[\x{2067}\x{2066}\x{2068}]+?(?-2)*)
    (?<pdf>([^\x{202B}\x{202A}\x{202E}\x{202D}]*)([^\x{202B}\x{202A}\x{202E}\x{202D}\x{202C}]*)
            ((?-2)[\x{202B}\x{202A}\x{202E}\x{202D}](?-2)(?-1)*(?-2)[\x{202C}](?-2))*
            (?-3)[\x{202B}\x{202A}\x{202E}\x{202D}]+?(?-2)*)
    (?<unbal>(?&pdi)|(?&pdf))(?<string>(?:'(?&unbal)')|(?:"(?&unbal)"))
    (?<comment>(?:\/\*(?&unbal)\*\/)|(?:\/\/(?&unbal)$)|(?:#(?&unbal)$))
)
(?&string)|(?&comment)
\end{lstlisting}
\caption{Regular Expression in PCRE2 syntax for identifying unbalanced Bidi control characters in comments and strings that may indicate Trojan Source attacks. Newlines added for formatting purposes.}
\label{fig:regex}
\end{figure*}

%% file: main.bbl
\begin{thebibliography}{10}
\providecommand{\url}[1]{#1}
\csname url@samestyle\endcsname
\providecommand{\newblock}{\relax}
\providecommand{\bibinfo}[2]{#2}
\providecommand{\BIBentrySTDinterwordspacing}{\spaceskip=0pt\relax}
\providecommand{\BIBentryALTinterwordstretchfactor}{4}
\providecommand{\BIBentryALTinterwordspacing}{\spaceskip=\fontdimen2\font plus
\BIBentryALTinterwordstretchfactor\fontdimen3\font minus
  \fontdimen4\font\relax}
\providecommand{\BIBforeignlanguage}[2]{{%
\expandafter\ifx\csname l@#1\endcsname\relax
\typeout{** WARNING: IEEEtran.bst: No hyphenation pattern has been}%
\typeout{** loaded for the language `#1'. Using the pattern for}%
\typeout{** the default language instead.}%
\else
\language=\csname l@#1\endcsname
\fi
#2}}
\providecommand{\BIBdecl}{\relax}
\BIBdecl

\bibitem{DBLP:journals/cacm/Thompson84}
\BIBentryALTinterwordspacing
K.~Thompson, ``Reflections on {Trusting} {Trust},'' \emph{Commun. {ACM}},
  vol.~27, no.~8, pp. 761--763, 1984. [Online]. Available:
  \url{https://doi.org/10.1145/358198.358210}
\BIBentrySTDinterwordspacing

\bibitem{9382367}
S.~Peisert, B.~Schneier, H.~Okhravi, F.~Massacci, T.~Benzel, C.~Landwehr,
  M.~Mannan, J.~Mirkovic, A.~Prakash, and J.~Michael, ``Perspectives on the
  {SolarWinds} {Incident},'' \emph{IEEE Security \& Privacy}, vol.~19, no.~02,
  pp. 7--13, Mar 2021.

\bibitem{unicode_bidi_2020}
\BIBentryALTinterwordspacing
{The Unicode Consortium}, ``\BIBforeignlanguage{en}{Unicode {Bidirectional}
  {Algorithm}},'' The Unicode Consortium, Tech. Rep. Unicode Technical Report
  \#9, Feb. 2020. [Online]. Available:
  \url{https://www.unicode.org/reports/tr9/tr9-42.html}
\BIBentrySTDinterwordspacing

\bibitem{painter_correctness_1967}
\BIBentryALTinterwordspacing
J.~Painter and J.~McCarthy, ``Correctness of a compiler for arithmetic
  expressions,'' in \emph{Proceedings of {Symposia} in {Applied}
  {Mathematics}}, vol.~19.\hskip 1em plus 0.5em minus 0.4em\relax American
  Mathematical Society, 1967, pp. 33--41. [Online]. Available:
  \url{http://jmc.stanford.edu/articles/mcpain/mcpain.pdf}
\BIBentrySTDinterwordspacing

\bibitem{dave2003compiler}
M.~A. Dave, ``Compiler verification: a bibliography,'' \emph{ACM SIGSOFT
  Software Engineering Notes}, vol.~28, no.~6, pp. 2--2, 2003.

\bibitem{patterson2019next}
D.~Patterson and A.~Ahmed, ``The next 700 compiler correctness theorems
  (functional pearl),'' \emph{Proceedings of the ACM on Programming Languages},
  vol.~3, no. ICFP, pp. 1--29, 2019.

\bibitem{7163211}
V.~D'Silva, M.~Payer, and D.~Song, ``The {Correctness}-{Security} {Gap} in
  {Compiler} {Optimization},'' in \emph{2015 IEEE Security and Privacy
  Workshops}, 2015, pp. 73--87.

\bibitem{8406587}
L.~Simon, D.~Chisnall, and R.~Anderson, ``What {You} {Get} is {What} {You} {C}:
  {Controlling} {Side} {Effects} in {Mainstream} {C} {Compilers},'' in
  \emph{2018 IEEE European Symposium on Security and Privacy ({EuroS\&P})},
  Apr. 2018, pp. 1--15.

\bibitem{unicode_2020}
\BIBentryALTinterwordspacing
{The Unicode Consortium}, ``\BIBforeignlanguage{en}{The {Unicode} {Standard},
  {Version} 13.0},'' Mar. 2020. [Online]. Available:
  \url{https://www.unicode.org/versions/Unicode13.0.0}
\BIBentrySTDinterwordspacing

\bibitem{5718996}
C.~J. Alberts, A.~J. Dorofee, R.~Creel, R.~J. Ellison, and C.~Woody, ``A
  {Systemic} {Approach} for {Assessing} {Software} {Supply}-{Chain} {Risk},''
  in \emph{2011 44th Hawaii International Conference on System Sciences}, 2011,
  pp. 1--8.

\bibitem{7163055}
A.~Nappa, R.~Johnson, L.~Bilge, J.~Caballero, and T.~Dumitras, ``The {Attack}
  of the {Clones}: {A} {Study} of the {Impact} of {Shared} {Code} on
  {Vulnerability} {Patching},'' in \emph{2015 IEEE Symposium on Security and
  Privacy}, 2015, pp. 692--708.

\bibitem{jospeh_biden_jr_executive_2021}
\BIBentryALTinterwordspacing
J.~Biden, ``\BIBforeignlanguage{en-US}{Executive {Order} on {Improving} the
  {Nation}'s {Cybersecurity}},'' May 2021, {Executive} {Order} 14028. [Online].
  Available:
  \url{https://www.whitehouse.gov/briefing-room/presidential-actions/2021/05/12/executive-order-on-improving-the-nations-cybersecurity}
\BIBentrySTDinterwordspacing

\bibitem{5428501}
R.~J. Ellison and C.~Woody, ``Supply-{Chain} {Risk} {Management}:
  {Incorporating} {Security} into {Software} {Development},'' in \emph{2010
  43rd Hawaii International Conference on System Sciences}, 2010, pp. 1--10.

\bibitem{1203227}
E.~Levy, ``Poisoning the software supply chain,'' \emph{IEEE Security Privacy},
  vol.~1, no.~3, pp. 70--73, 2003.

\bibitem{7180277}
B.~A. Sabbagh and S.~Kowalski, ``A {Socio-technical} {Framework} for {Threat}
  {Modeling} a {Software} {Supply} {Chain},'' \emph{IEEE Security Privacy},
  vol.~13, no.~4, pp. 30--39, 2015.

\bibitem{10.1007/978-3-030-52683-2_2}
M.~Ohm, H.~Plate, A.~Sykosch, and M.~Meier, ``Backstabber's {Knife}
  {Collection}: {A} {Review} of {Open} {Source} {Software} {Supply} {Chain}
  {Attacks},'' in \emph{Detection of Intrusions and Malware, and Vulnerability
  Assessment}, C.~Maurice, L.~Bilge, G.~Stringhini, and N.~Neves, Eds.\hskip
  1em plus 0.5em minus 0.4em\relax Cham: Springer International Publishing,
  2020, pp. 23--43.

\bibitem{owasp_a92017}
\BIBentryALTinterwordspacing
OWASP, ``A9:2017 {Using} {Components} with {Known} {Vulnerabilities},'' 2017.
  [Online]. Available:
  \url{https://owasp.org/www-project-top-ten/2017/A9_2017-Using_Components_with_Known_Vulnerabilities.html}
\BIBentrySTDinterwordspacing

\bibitem{brian_krebs_right--left_2011}
\BIBentryALTinterwordspacing
{Brian Krebs}, ``\BIBforeignlanguage{en-US}{‘{Right}-to-{Left} {Override}’
  {Aids} {Email} {Attacks}},'' Sep. 2011. [Online]. Available:
  \url{https://krebsonsecurity.com/2011/09/right-to-left-override-aids-email-attacks/}
\BIBentrySTDinterwordspacing

\bibitem{boucher2021bad}
N.~Boucher, I.~Shumailov, R.~Anderson, and N.~Papernot, ``Bad {Characters}:
  {Imperceptible} {NLP} {Attacks},'' in \emph{43rd IEEE Symposium on Security
  and Privacy}.\hskip 1em plus 0.5em minus 0.4em\relax IEEE, 2022.

\bibitem{7546508}
Y.~Acar, M.~Backes, S.~Fahl, D.~Kim, M.~L. Mazurek, and C.~Stransky, ``You
  {Get} {Where} {You're} {Looking} for: {The} {Impact} of {Information}
  {Sources} on {Code} {Security},'' in \emph{2016 IEEE Symposium on Security
  and Privacy (SP)}, 2016, pp. 289--305.

\bibitem{stackoverflow_android}
F.~Fischer, K.~Böttinger, H.~Xiao, C.~Stransky, Y.~Acar, M.~Backes, and
  S.~Fahl, ``{Stack Overflow Considered Harmful? The Impact of Copy\&Paste on
  Android Application Security},'' in \emph{2017 IEEE Symposium on Security and
  Privacy (SP)}, 2017, pp. 121--136.

\bibitem{stackoverflow_voting}
D.~van~der Linden, E.~Williams, J.~Hallett, and A.~Rashid, ``{The Impact of
  Surface Features on Choice of (in)Secure Answers by Stackoverflow Readers},''
  \emph{IEEE Transactions on Software Engineering}, vol.~48, no.~2, pp.
  364--376, 2022.

\bibitem{SimpsonMC20apwg}
G.~Simpson, T.~Moore, and R.~Clayton, ``Ten years of attacks on companies using
  visual impersonation of domain names,'' in \emph{{APWG} Symposium on
  Electronic Crime Research (eCrime)}.\hskip 1em plus 0.5em minus 0.4em\relax
  IEEE, 2020.

\bibitem{sullivan_paypal_2000}
\BIBentryALTinterwordspacing
B.~Sullivan, ``\BIBforeignlanguage{en}{{PayPal} alert! {Beware} the '{Paypai}'
  scam},'' Jul. 2000. [Online]. Available:
  \url{https://www.zdnet.com/article/paypal-alert-beware-the-paypai-scam-5000109103/}
\BIBentrySTDinterwordspacing

\bibitem{unicode_security_2014}
\BIBentryALTinterwordspacing
{The Unicode Consortium}, ``\BIBforeignlanguage{en}{Unicode {Security}
  {Considerations}},'' The Unicode Consortium, Tech. Rep. Unicode Technical
  Report \#36, Sep. 2014. [Online]. Available:
  \url{https://www.unicode.org/reports/tr36/tr36-15.html}
\BIBentrySTDinterwordspacing

\bibitem{10.1145/503124.503156}
\BIBentryALTinterwordspacing
E.~Gabrilovich and A.~Gontmakher, ``The {Homograph} {Attack},'' \emph{Commun.
  ACM}, vol.~45, no.~2, p. 128, Feb. 2002. [Online]. Available:
  \url{https://doi.org/10.1145/503124.503156}
\BIBentrySTDinterwordspacing

\bibitem{10.5555/1267359.1267383}
T.~Holgers, D.~E. Watson, and S.~D. Gribble, ``Cutting through the {Confusion}:
  {A} {Measurement} {Study} of {Homograph} {Attacks},'' in \emph{Proceedings of
  the Annual Conference on USENIX '06 Annual Technical Conference}, ser. ATEC
  '06.\hskip 1em plus 0.5em minus 0.4em\relax USA: USENIX Association, 2006,
  p.~24.

\bibitem{noauthor_capec-632_2015}
\BIBentryALTinterwordspacing
MITRE, ``{CAPEC}-632: {Homograph} {Attack} via {Homoglyphs} ({Version} 3.4),''
  MITRE, Common {Attack} {Pattern} {Enumeration} and {Classification} 632, Nov.
  2015. [Online]. Available:
  \url{https://capec.mitre.org/data/definitions/632.html}
\BIBentrySTDinterwordspacing

\bibitem{10.1145/3355369.3355587}
\BIBentryALTinterwordspacing
H.~Suzuki, D.~Chiba, Y.~Yoneya, T.~Mori, and S.~Goto, ``{ShamFinder}: {An}
  {Automated} {Framework} for {Detecting} {IDN} {Homographs},'' in
  \emph{Proceedings of the Internet Measurement Conference}, ser. IMC
  '19.\hskip 1em plus 0.5em minus 0.4em\relax New York, NY, USA: Association
  for Computing Machinery, 2019, p. 449–462. [Online]. Available:
  \url{https://doi.org/10.1145/3355369.3355587}
\BIBentrySTDinterwordspacing

\bibitem{RFC3492}
\BIBentryALTinterwordspacing
A.~Costello, ``{Punycode}: {A} {Bootstring} encoding of {Unicode} for
  {Internationalized} {Domain} {Names} in {Applications} {(IDNA)},'' Internet
  Requests for Comments, RFC 3492, March 2003. [Online]. Available:
  \url{https://www.rfc-editor.org/rfc/rfc3492}
\BIBentrySTDinterwordspacing

\bibitem{RFC5891}
\BIBentryALTinterwordspacing
J.~Klensin, ``{Internationalized} {Domain} {Names} in {Applications} {(IDNA)}:
  {Protocol},'' Internet Requests for Comments, RFC 5891, August 2010.
  [Online]. Available: \url{https://www.rfc-editor.org/rfc/rfc5891}
\BIBentrySTDinterwordspacing

\bibitem{microsoft_win32sirefef_2017}
\BIBentryALTinterwordspacing
{Microsoft}, ``Win32/{Sirefef},'' Sep. 2017. [Online]. Available:
  \url{https://www.microsoft.com/en-us/wdsi/threats/malware-encyclopedia-description?Name=Win32/Sirefef}
\BIBentrySTDinterwordspacing

\bibitem{jakob_lell_hacking-contest_2014}
\BIBentryALTinterwordspacing
J.~Lell, ``\BIBforeignlanguage{en-US}{[{Hacking}-{Contest}] {Invisible}
  configuration file backdooring with {Unicode} homoglyphs},'' May 2014.
  [Online]. Available:
  \url{https://www.jakoblell.com/blog/2014/05/07/hacking-contest-invisible-configuration-file-backdooring-with-unicode-homoglyphs/}
\BIBentrySTDinterwordspacing

\bibitem{wheeler_2020_underhanded}
\BIBentryALTinterwordspacing
D.~A. Wheeler, ``Initial analysis of underhanded source code,'' Institute for
  Defense Analyses, Tech. Rep. D-13166, 2020. [Online]. Available:
  \url{https://apps.dtic.mil/sti/pdfs/AD1122149.pdf}
\BIBentrySTDinterwordspacing

\bibitem{nistvulntaxonomy}
\BIBentryALTinterwordspacing
``{A Taxonomy of Software Flaws},'' May 2021. [Online]. Available:
  \url{https://www.nist.gov/itl/ssd/software-quality-group/taxonomy-software-flaws}
\BIBentrySTDinterwordspacing

\bibitem{mitre_cve_2021}
\BIBentryALTinterwordspacing
{MITRE}, ``About the {CVE} {Program},'' Oct. 2021. [Online]. Available:
  \url{https://www.cve.org/About/Overview}
\BIBentrySTDinterwordspacing

\bibitem{solardesigner1997}
\BIBentryALTinterwordspacing
{Solar Designer}, ``Getting around non-executable stack (and fix),'' Aug 1997.
  [Online]. Available: \url{https://seclists.org/bugtraq/1997/Aug/63}
\BIBentrySTDinterwordspacing

\bibitem{10.1145/2133375.2133377}
\BIBentryALTinterwordspacing
R.~Roemer, E.~Buchanan, H.~Shacham, and S.~Savage, ``{Return-Oriented
  Programming: Systems, Languages, and Applications},'' \emph{ACM Trans. Inf.
  Syst. Secur.}, vol.~15, no.~1, mar 2012. [Online]. Available:
  \url{https://doi.org/10.1145/2133375.2133377}
\BIBentrySTDinterwordspacing

\bibitem{ISO:2018:III}
\BIBentryALTinterwordspacing
{ISO}, \emph{{{ISO}\slash{}{IEC} 9899:2018 {Information} technology ---
  {Programming} languages --- {C}}}, 4th~ed.\hskip 1em plus 0.5em minus
  0.4em\relax Geneva, Switzerland: International Organization for
  Standardization, Jun. 2018. [Online]. Available:
  \url{https://www.iso.org/standard/74528.html}
\BIBentrySTDinterwordspacing

\bibitem{ISO:2020:IIIa}
\BIBentryALTinterwordspacing
ISO, \emph{{{ISO}\slash{}{IEC} 14882:2020 {Information} technology ---
  {Programming} languages --- {C++}}}, 6th~ed.\hskip 1em plus 0.5em minus
  0.4em\relax Geneva, Switzerland: International Organization for
  Standardization, Dec. 2020. [Online]. Available:
  \url{https://www.iso.org/standard/79358.html}
\BIBentrySTDinterwordspacing

\bibitem{ISO:2018:CS}
\BIBentryALTinterwordspacing
{ISO}, \emph{{{ISO}\slash{}{IEC} 23270:2018 {Information} technology ---
  {Programming} languages --- {C\#}}}, 3rd~ed.\hskip 1em plus 0.5em minus
  0.4em\relax Geneva, Switzerland: International Organization for
  Standardization, Dec. 2018. [Online]. Available:
  \url{https://www.iso.org/standard/75178.html}
\BIBentrySTDinterwordspacing

\bibitem{Ecma:2021}
\BIBentryALTinterwordspacing
Ecma, \emph{{ECMA-262}}, 12th~ed.\hskip 1em plus 0.5em minus 0.4em\relax
  Geneva, Switzerland: {Ecma} {International}, Jun. 2021. [Online]. Available:
  \url{https://www.ecma-international.org/publications-and-standards/standards/ecma-262}
\BIBentrySTDinterwordspacing

\bibitem{gosling_java_2021}
\BIBentryALTinterwordspacing
J.~Gosling, B.~Joy, G.~Steele, G.~Bracha, A.~Buckley, D.~Smith, and G.~Bierman,
  \emph{\BIBforeignlanguage{en}{The {Java}® {Language} {Specification}}},
  16th~ed.\hskip 1em plus 0.5em minus 0.4em\relax {Java} {Community} {Press},
  Feb. 2021. [Online]. Available:
  \url{https://docs.oracle.com/javase/specs/jls/se16/jls16.pdf}
\BIBentrySTDinterwordspacing

\bibitem{rustref}
\BIBentryALTinterwordspacing
{The Rust Project Developers}, \emph{The {Rust} {Reference}}.\hskip 1em plus
  0.5em minus 0.4em\relax {The} {Rust} {Foundation}, 2018. [Online]. Available:
  \url{https://doc.rust-lang.org/reference}
\BIBentrySTDinterwordspacing

\bibitem{goref}
\BIBentryALTinterwordspacing
{The Go Project Developers}, \emph{The {Go} {Programming} {Language}
  {Specification}}.\hskip 1em plus 0.5em minus 0.4em\relax Google, Feb. 2021.
  [Online]. Available: \url{https://golang.org/ref/spec}
\BIBentrySTDinterwordspacing

\bibitem{pythonref}
\BIBentryALTinterwordspacing
{The Python Project Developers}, \emph{{The} {Python} {Language} {Reference}},
  3rd~ed.\hskip 1em plus 0.5em minus 0.4em\relax {The} {Python} {Software}
  {Foundation}, 2018. [Online]. Available:
  \url{https://docs.python.org/3/reference}
\BIBentrySTDinterwordspacing

\bibitem{lwnbackdoor}
\BIBentryALTinterwordspacing
J.~Corbet, ``An attempt to backdoor the kernel,'' \emph{Linux Weekly News},
  Nov. 2003. [Online]. Available: \url{https://lwn.net/Articles/57135}
\BIBentrySTDinterwordspacing

\bibitem{7958574}
F.~Fischer, K.~Böttinger, H.~Xiao, C.~Stransky, Y.~Acar, M.~Backes, and
  S.~Fahl, ``Stack {Overflow} {Considered} {Harmful}? {The} {Impact} of {Copy}
  \& {Paste} on {Android} {Application} {Security},'' in \emph{2017 IEEE
  Symposium on Security and Privacy (SP)}, 2017, pp. 121--136.

\bibitem{Perlroth2021}
N.~Perlroth, \emph{This {Is} {How} {They} {Tell} {Me} the {World} {Ends}: {The}
  {Cyberweapons} {Arms} {Race}}.\hskip 1em plus 0.5em minus 0.4em\relax
  Bloomsbury, 2021.

\bibitem{mitre_cwe_1007}
\BIBentryALTinterwordspacing
{MITRE}, ``{CWE} 1007: {Insufficient} {Visual} {Distinction} of {Homoglyphs}
  {Presented} to {User},'' Jul. 2017. [Online]. Available:
  \url{https://cwe.mitre.org/data/definitions/1007.html}
\BIBentrySTDinterwordspacing

\bibitem{feist_slither_2018}
\BIBentryALTinterwordspacing
J.~Feist, ``\BIBforeignlanguage{en-US}{Slither – a {Solidity} static analysis
  framework},'' Oct. 2018. [Online]. Available:
  \url{https://blog.trailofbits.com/2018/10/19/slither-a-solidity-static-analysis-framework/}
\BIBentrySTDinterwordspacing

\bibitem{github_issue_go}
\BIBentryALTinterwordspacing
{Golang project contributors}, ``proposal: spec: disallow ltr/rtl characters in
  string literals?'' May 2017. [Online]. Available:
  \url{https://github.com/golang/go/issues/20209}
\BIBentrySTDinterwordspacing

\bibitem{certcc}
\BIBentryALTinterwordspacing
{Carnegie Mellon University Software Engineering Institute}, ``{CERT
  Coordination Center}.'' [Online]. Available: \url{https://www.kb.cert.org}
\BIBentrySTDinterwordspacing

\bibitem{distros_list}
\BIBentryALTinterwordspacing
{Openwall Project}, ``Operating {System} {Distribution} {Security} {Contact}
  lists,'' Sep 2021. [Online]. Available:
  \url{https://oss-security.openwall.org/wiki/mailing-lists/distros}
\BIBentrySTDinterwordspacing

\bibitem{mitre_cwe_2021}
\BIBentryALTinterwordspacing
{MITRE}, ``{CWE} {Overview},'' Oct. 2021. [Online]. Available:
  \url{https://cwe.mitre.org/about/index.html}
\BIBentrySTDinterwordspacing

\bibitem{github_advisory}
\BIBentryALTinterwordspacing
{GitHub}, ``Warning about bidirectional {Unicode} text,'' Oct. 2021. [Online].
  Available:
  \url{https://github.blog/changelog/2021-10-31-warning-about-bidirectional-unicode-text}
\BIBentrySTDinterwordspacing

\bibitem{atlassian_advisory}
\BIBentryALTinterwordspacing
{Atlassian}, ``{Multiple Products Security Advisory - Unrendered unicode
  bidirectional override characters - CVE-2021-42574},'' Nov. 2021. [Online].
  Available:
  \url{https://confluence.atlassian.com/security/multiple-products-security-advisory-unrendered-unicode-bidirectional-override-characters-cve-2021-42574-1086419475.html}
\BIBentrySTDinterwordspacing

\bibitem{gitlab_advisory}
\BIBentryALTinterwordspacing
{GitLab}, ``{GitLab Security Release: 14.4.1, 14.3.4, and 14.2.6},'' Oct. 2021.
  [Online]. Available:
  \url{https://about.gitlab.com/releases/2021/10/28/security-release-gitlab-14-4-1-released}
\BIBentrySTDinterwordspacing

\bibitem{vscode_advisory}
\BIBentryALTinterwordspacing
{Microsoft}, ``{Visual Studio Code: October 2021 (version 1.62)},'' Oct. 2021.
  [Online]. Available: \url{https://code.visualstudio.com/updates/v1_62}
\BIBentrySTDinterwordspacing

\bibitem{emacs_patch}
\BIBentryALTinterwordspacing
E.~Zaretskii, ``{Better detection of potentially malicious bidi text},'' Nov.
  2021. [Online]. Available:
  \url{https://git.savannah.gnu.org/cgit/emacs.git/commit/?id=b96855310efed13e0db1403759b686b9bc3e7490}
\BIBentrySTDinterwordspacing

\bibitem{rust_advisory}
\BIBentryALTinterwordspacing
{The Rust Security Response WG}, ``{Security advisory for rustc
  (CVE-2021-42574)},'' Nov. 2021. [Online]. Available:
  \url{https://blog.rust-lang.org/2021/11/01/cve-2021-42574.html}
\BIBentrySTDinterwordspacing

\bibitem{gcc_wbidichars}
\BIBentryALTinterwordspacing
\vspace{0mm}GNU, ``{GCC: Warning Options},'' Jan. 2022. [Online]. Available:
  \url{https://gcc.gnu.org/onlinedocs/gcc/Warning-Options.html}
\BIBentrySTDinterwordspacing

\bibitem{llvm_advisory}
\BIBentryALTinterwordspacing
LLVM, ``{New passes in clang-tidy to detect (some) Trojan Source},'' Jan. 2021.
  [Online]. Available:
  \url{https://blog.llvm.org/posts/2022-01-12-trojan-source}
\BIBentrySTDinterwordspacing

\bibitem{unicode_avoiding_spoofing}
\BIBentryALTinterwordspacing
M.~Davis, R.~Leroy, P.~Constable, and M.~Scherer, ``{Avoiding Source Code
  Spoofing},'' Jan. 2022. [Online]. Available:
  \url{https://www.unicode.org/L2/L2022/22007-avoiding-spoof.pdf}
\BIBentrySTDinterwordspacing

\end{thebibliography}
